\documentclass[aps,prb,reprint,amsmath,superscriptaddress,floatfix]{revtex4-2}
\usepackage{bm}
\usepackage{amssymb}
\usepackage{graphicx}
\usepackage[usenames]{color}

\newcommand{\LiHoF}{LiHoF$_4$}
\newcommand{\lhf}{LiHoF$_4$}

\begin{document}

\title{A Gallery of Soft Modes: Theory and Experiment at a Ferromagnetic Quantum Phase Transition}

\author{P.C.E. Stamp}
\thanks{These authors contributed equally to this work.}
\affiliation{Department of Physics and Astronomy, University of British Columbia,
Vancouver, British Columbia V6T 1Z1, Canada}
\affiliation{Pacific Institute of Theoretical Physics, University of British
Columbia, Vancouver, British Columbia V6T 1Z1, Canada}
\author{D.M. Silevitch}
\thanks{These authors contributed equally to this work.}
\affiliation{Division of Physics, Mathematics, and Astronomy, California Institute
of Technology, Pasadena California 91125, USA}
\author{M. Libersky}
\affiliation{Division of Physics, Mathematics, and Astronomy, California Institute
of Technology, Pasadena California 91125, USA}
\author{Ryan McKenzie}
        \affiliation{Ames National Laboratory, U.S. DOE, Iowa State University, Ames, Iowa 50011, USA}
        \affiliation{Department of Physics and Astronomy, Iowa State University, Ames, Iowa 50011, USA}
\author{A. A. Geim}
\affiliation{Department of Physics, Harvard University, Cambridge Massachusetts
02138, USA}
\author{T.F. Rosenbaum}
\email[Correspondence and requests for materials should be addressed to
T.F.R.,]{tfr@caltech.edu}
\affiliation{Division of Physics, Mathematics, and Astronomy, California Institute
of Technology, Pasadena California 91125, USA}
\date{\today}

\begin{abstract}
We examine the low-energy excitations in the vicinity of the quantum critical point in \LiHoF, a physical realization of the Transverse Field Ising Model, focusing on the long-range fluctuations which soften to zero energy at the ferromagnetic quantum phase transition. Microwave spectroscopy in tunable loop-gap resonator structures identifies and characterizes the soft mode and higher-energy electronuclear states. We study these modes as a function of frequency and magnetic fields applied transverse and parallel to the Ising axis. These are understood in the context of a theoretical model of a soft electronuclear mode that interacts with soft photons as well as soft phonons. We identify competing infrared divergences at the quantum critical point, coming from the photons and the electronuclear soft mode. It is an incomplete cancellation of these divergences that leads to the muted but distinct signatures observed in the experiments. The application of a longitudinal magnetic field gaps the soft mode. Measurements well away from the quantum critical point reveal a set of ``Walker'' modes associated with ferromagnetic domain dynamics.
\end{abstract}

\maketitle


\section{Introduction}


At the approach to a quantum critical point, long-wavelength fluctuations grow and their energy dives to zero. This ``soft mode'' has been a long-accepted paradigm for quantum phase transitions \cite{hertzQuantumCriticalPhenomena1976,belitzHowGenericScale2005}, but only recently has been measured directly \cite{LiberskySM}. Experiments on a physical realization of the canonical quantum spin model, the Ising magnet in transverse magnetic field, demonstrated that the soft mode is robust even in the presence of potential sources of disruption from crystalline defects and a fluctuating nuclear spin bath. This gapless behavior at the ferromagnetic quantum critical point in \LiHoF\ was not observed in earlier neutron scattering measurements \cite{ronnowQuantumPhaseTransition2005}, which probed excitations matched to the splitting between electronic spin levels. Instead, microwave measurements at milliKelvin temperatures were required to track the softening of the lowest electronuclear mode, with level spacings $\sim$GHz (0.05 K).

The signature of the divergent mode recorded from the microwave resonator as the transverse magnetic field was swept through the critical point was surprisingly weak \cite{LiberskySM}. We show here that an understanding of the magnon-photon interactions in the resonator is necessary to account for the experimental results. Crucially, competing and canceling infrared divergences, familiar from quantum electrodynamics and condensed matter problems like the Kondo effect, result in the soft mode emerging as a weak avoided level crossing. In this paper, we develop a theory of magnon-photon hybridization to motivate the experimental results, as well as an explicit treatment of resonator performance, the magnetic modes in the quantum magnet, and magnetic domain physics germane to the sample response.

The electronuclear soft mode and the photons are not the only soft modes in the system. There are soft acoustic phonons, which also interact with the electronuclear mode, and we highlight their main features. The textbook description of a single soft mode at \LiHoF's quantum critical point simply does not capture the experimental reality of multiple soft modes in interaction with each other, all influencing the onset of magnetic order at the quantum phase transition.

More generally, our results connect to studies in recent years of optical crystals doped with rare earth ions as a mechanism for microwave-optical conversion in hybrid quantum systems \cite{bartholomewOnchipCoherentMicrowavetooptical2020a,rochmanMicrowavetoopticalTransductionErbium2023a,serranoUltranarrowOpticalLinewidths2022}. Similarly, magnetic crystals such as YIG provide insight into fundamental physical phenomena such as Bose-Einstein condensation \cite{bozhkoSupercurrentRoomtemperatureBose2016} and information processing using spin excitations \cite{liHybridMagnonicsPhysics2020}. Fully concentrated rare earth ion crystals can combine these properties, with a range of magnetic ordering properties \cite{evertsMicrowaveOpticalPhoton2019} and implications for quantum information processing.

We focus here on perhaps the best-understood example of such a system, Li(RE,Y)F$_4$, where different rare earths (RE) result in different magnetic ground states with different symmetries, and the possibility of chemical dilution by nonmagnetic Y$^{3+}$. Large single crystals are commercially available due to the suitability of this family for infrared lasing rods \cite{chicklisHighEfficiencyRoom1971}. The pure compound, with Ho$^{3+}$ as the sole rare earth, is a dipolar-coupled ferromagnet with Ising spin symmetry and a Curie temperature of 1.53 K.

The remainder of this paper is structured as follows. In Section~\ref{sec:ENmodes}, we describe the microscopic Hamiltonian and resultant energy mode structure for \LiHoF, followed in Section~\ref{sec:resonator} by a discussion of the microwave resonator techniques used to  probe this mode structure. In Section~\ref{sec:MPhybridization}, we discuss the implications of the combined \LiHoF\  + resonator system, detailing the magnon-photon hybridization that results. In Section~\ref{sec:SoftPhotons}, we develop the theory of mode softening at the quantum critical point in the hybridized system, including the essential result of a cancellation of infrared divergences leading to a weak but non-zero physical signature of the soft mode. Having developed this model, we tie it back to the measured data in Sections~\ref{sec:Spectra} and \ref{sec:long_field}, and show that that the soft mode and related effects at the quantum phase transition (QPT) can be detected in the proper experimental geometry. Finally, in Section~\ref{sec:domains}, we show some additional features in the data arising from effects unrelated to quantum criticality, and in particular discuss the effects of domain dynamics in the microwave response.


\section{Hamiltonian and Electronuclear Modes}
\label{sec:ENmodes}


\lhf\ is the canonical physical realization of the transverse field Ising model \cite{Hansen:1975p53,Bitko:1996p44}. The Ho$^{3+}$ ions form a tetragonal scheelite structure with hyperfine coupling to the $^{165}$Ho nuclei (holmium is monoisotopic). The Hamiltonian can be written as \cite{chakrabortyTheoryMagneticPhase2004}:

\begin{multline}
    \mathcal{H} = \sum_i V_C\left(\vec{J}_i\right) -g_L\mu_B\sum_i B_x J_i^x
    +A\sum_i\vec{I}_i\cdot\vec{J}_i  \\ - \frac{1}{2} J_D \sum_{i\neq j}
    D_{ij}^{\mu\nu} J_i^\mu J_j^\nu +\frac{1}{2} J_{nn}\sum_{\langle ij \rangle}
    \vec{J}_i \cdot \vec{J}_j.
\end{multline}
The crystal field, transverse field Zeeman, hyperfine, dipolar coupling, and antiferromagnetic exchange couplings comprise the successive terms in the Hamiltonian.

\begin{figure}
    \includegraphics[width=2.75in]{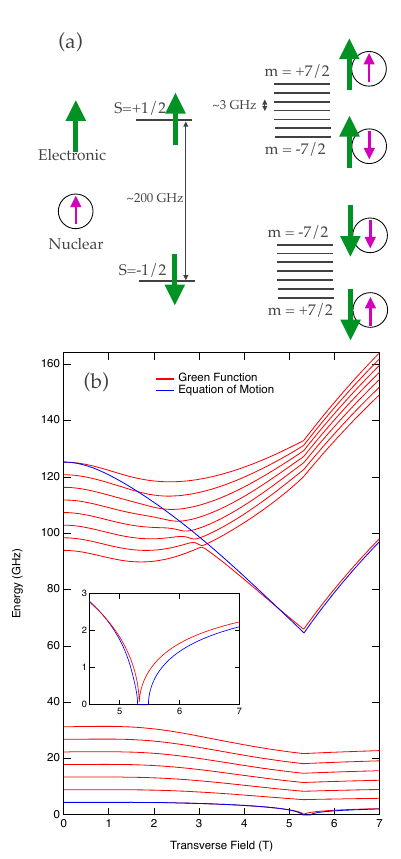}
\caption{Energy level structure of \LiHoF. (a) Zero-field energy level hierarchy. Taken alone, the electronic spins form $S=1/2$ doublets with an energy gap of order 200 GHz. Hybridization with the holmium nuclear spins gives rise to a ladder of states spaced approximately 3 GHz apart  (b) Finite-transverse-field electronuclear modes calculated in the random phase approximation (RPA) using a Green's function approach (red) or the equation of motion of the magnetization (blue) (see text). The mode that splits away from the upper band near 3 T is the dominant mode in neutron scattering.  Green's function curves adapted from Ref. \cite{mckenzieThermodynamicsQuantumIsing2018}.}
\label{fig:energy}
\end{figure}

These interactions result in the general energy level structure shown in Fig.~\ref{fig:energy}(a). At low temperatures ($T\sim 1\:$K) the crystal field ground state doublet provides an effective spin $1/2$ electronic spin, with these two lowest-lying levels split into eight hyperfine levels by the $^{165}$Ho nuclei. One may truncate the Hamiltonian down to the two lowest electronic levels leading to an effective Hamiltonian in which the Ising nature of the material is apparent \cite{chakrabortyTheoryMagneticPhase2004,Tabei,mckenzieThermodynamicsQuantumIsing2018}:
\begin{align}
\label{eq:LiHoTrunc}
\mathcal{H} =
- \frac{1}{2} \sum_{i \neq j} V_{ij}^{zz} \tau_{i}^{z}\tau_{j}^{z}
- \frac{\Delta}{2}\sum_{i}\tau_{i}^{x}
+ \mathcal{H}_{hyp}.
\end{align}
The truncated electronic spin operators are given in terms of Pauli operators by $J_i^{\mu} = C_{\mu}(B_x) + \sum_{\nu=x,y,z} C_{\mu\nu}(B_{x}) \tau_i^{\nu}$, and the effective transverse field $\Delta(B_x)$ is the splitting of the two lowest electronic levels in the physical transverse field $B_x$.

Solving the Hamiltonian at the mean field (MF) level provides good results given the long-range nature of the dipole interaction and the absence of conduction electrons. An exception is the lowest electronuclear mode, which remains gapped at the QPT. Adding fluctuations, but neglecting their interactions, leads to the random phase approximation (RPA) which predicts that the soft mode indeed goes to zero at the QPT (Fig.~\ref{fig:energy}(b) bottom right). This pair of plots also exhibits the $\sim3$~GHz energy scale difference between hyperfine levels. The mode that branches off from the higher cluster in the upper figure is the dominant mode seen in neutron scattering measurements \cite{ronnowQuantumPhaseTransition2005}. Exploring the structure of the lower energy dynamics requires alternative experimental methods, which we describe in Section \ref{sec:resonator} below.

One may calculate the electronuclear modes of the \lhf\ system making use of the Green's function, or alternatively, from the equation of motion for the system's spins. In the Green's function approach, one calculates the connected imaginary time Green's function, which, in Matsubara frequency space, is given by \cite{JensenMackintosh, mckenzieThermodynamicsQuantumIsing2018}
\begin{align}
\label{eq:GF}
G_{ij}^{\mu \nu}(i\omega_n) &= \frac{1}{\beta} \int_0^{\beta} d\tau\ G_{ij}^{\mu
\nu}(\tau)
\\ \nonumber
&=- \frac{1}{\beta} \int_0^{\beta} d\tau \
\langle T_{\tau}\ \delta J_i^{\mu}(\tau) \delta J_j^{\nu}(0)\rangle.
\end{align}
The poles of this function determine the modes of the system, and their residues determine the spectral weights. The dynamic susceptibility of the electronic spins follows from $\chi_J^{\mu \nu}(ij,\omega) = -\beta G_{ij}^{\mu \nu}(i\omega_n \rightarrow \omega + i0^+) = \widetilde{\chi}_J^{\mu \nu}(ij,\omega) + \chi_{J,el}^{\mu \nu}(ij) \delta_{\omega,0}$, where in the final expression the dynamic susceptibility is divided into an inelastic component plus an additional elastic term describing a quasi-elastic diffusive pole of the system. The quasi-elastic diffusive pole vanishes in the paramagnetic phase of the system and decays exponentially with temperature; we will neglect this component of the susceptibility in what follows.

Specializing to $\chi^{zz}$ and transforming to momentum space, the spectral representation of the inelastic component of the dynamic susceptibility is
\begin{align}
\label{eq:ChiSpin}
\widetilde{\chi}_J(\boldsymbol{k}, \omega) =  \sum_m
\biggr[\frac{A_{\boldsymbol{k}}^m 2E_{\boldsymbol{k}}^m}
{(E_{\boldsymbol{k}}^m)^2-(\omega+i\Gamma_m/2)^2} \biggr],
\end{align}
where $A_{\boldsymbol{k}}^m$ is the spectral weight of the $m^{th}$ mode $E_{\boldsymbol{k}}^m$ determined in the RPA \cite{JensenMackintosh,mckenzieTheoryMagnonPolaritons2022}. For the 16 level truncated LiHoF$_4$ Hamiltonian, given by equation (\ref{eq:LiHoTrunc}), there are 120 possible RPA modes corresponding to ground state and excited state transitions in the system. Note that the dynamic susceptibility of the spins has units of inverse energy; it is related to the (dimensionless) susceptibility of the material by $\chi = 4\pi \rho_s J_D \chi_J$, where $\rho_s$ is the spin density and $J_D = \mu_0 (g \mu_B)^2/(4\pi)$ \cite{JensenMackintosh}.  We will make use of the dynamic susceptibility to analyze experimental resonator transmission spectra in Section \ref{sec:Spectra}.

Alternatively, one may calculate the modes of the electronuclear system from the lossless Landau-Lifshitz equation $\dot{\boldsymbol{M}} = \gamma \ \boldsymbol{M} \times \boldsymbol{B}$, or, equivalently, the Heisenberg equation of motion of the spin operators. This approach is useful for analyzing magnetostatic ``Walker'' modes (see Section \ref{sec:domains}). However, it does not capture the excited state transitions determined by the dynamic susceptibility. The magnetization of the holmium ions in LiHoF$_4$ is given by $\boldsymbol{M}=\rho_s \boldsymbol{\mu}$, where $\rho_s = N/V$ is the density of the holmium ions and the magnetic moment at a particular site is given by $\boldsymbol{\mu} = \gamma_J \boldsymbol{J} + \gamma_I \boldsymbol{I}$. The ratio of gyromagnetic ratios is $|\gamma_I/\gamma_J| = 5.2 \times 10^{-4}$, so we will drop the nuclear contributions to the moment from subsequent calculations. Due to the strong hyperfine coupling, the time evolution of the electronic component of the moment is still strongly dependent on the nuclear spins. Rather than the Landau-Lifshitz equation, one may equivalently formulate the spin dynamics in terms of the Heisenberg equations of motion of the spin operators. In order to determine the electronuclear modes of LiHoF$_4$, we consider the operators $\boldsymbol{X}_i = \{\boldsymbol{\tau}_i, \boldsymbol{I}_i\}$, where in the truncated Hamiltonian the electronic spin operators are $J_i^{\mu} = C_{\mu}(B_x) + \sum_{\nu=x,y,z} C_{\mu\nu}(B_{x}) \tau_i^{\nu}$. The equation of motion is then
\begin{align}
\label{eq:motion}
\frac{d}{dt}\boldsymbol{X}_i
= \frac{i}{\hbar}[\mathcal{H},\boldsymbol{X}_i] = M \boldsymbol{X}_i,
\end{align}
where $M$ is a $6 \times 6$ matrix governing the electronuclear spin dynamics. To proceed, we expand in fluctuations of the spins about their MFs, and treat the fluctuations in the RPA.

In the RPA \cite{Blinc}, we consider fluctuations of the electronic spins about their MF values $ \langle \tau_i^{\mu}(t) \rangle  = \langle \tau^{\mu} \rangle_0 + \langle \delta \tau_i^{\mu}(t) \rangle$, and likewise for the nuclear spins. The interaction between spins at different sites is decoupled, so that $\langle \tau_i^{\mu} \tau_j^{\nu}\rangle = \langle \tau_i^{\mu} \rangle\langle \tau_j^{\nu}\rangle$ for any $i \neq j$. Furthermore, we decouple the interaction between the electronic and nuclear spins $\langle \tau_i^{\mu} I_i^{\nu}\rangle = \langle \tau_i^{\mu} \rangle\langle I_i^{\nu}\rangle$. This step is unnecessary in the Green's function approach, in which we work in a larger Hilbert space consisting of the single ion electronuclear Hubbard operators. The equation of motion in the RPA becomes
\begin{align}
\label{eq:dynamics}
\frac{d}{dt}\delta \boldsymbol{X}_i = M_\mathrm{RPA} \delta \boldsymbol{X}_i
\quad \text{and} \quad
M_{MF} \langle \boldsymbol{X} \rangle_0=0,
\end{align}
The eigenvalues of $M_\mathrm{RPA}$ determine the electronuclear modes of the LiHoF$_4$ system and the equation on the right determines the MF polarizations of the electronic and nuclear spins. The RPA decoupling of the electronic and nuclear spins leads to a discrepancy in the spin polarizations determined by the MF component of equation (\ref{eq:dynamics}), and the spin polarizations determined self consistently from the MF component of equation (\ref{eq:LiHoTrunc}). In what follows, we will use the MF spin polarizations determined self-consistently from equation (\ref{eq:LiHoTrunc}), rather than the values determined from the dynamic equation.

In the absence of the hyperfine interaction, the electronic equation of motion yields a ``longitudinal'' zero mode, $\omega_{\parallel}=0$, as well as a pair of ``transverse'' modes, $\pm \omega_{\perp}$, where ``longitudinal'' and ``transverse''  mean parallel and orthogonal to the spin polarization, respectively. When the nuclear spins are included, the RPA equation of motion yields two zero modes, and a pair of electronuclear modes at positive and negative frequencies. The upper electronuclear mode corresponds to the electronic excitation that has been measured in neutron scattering experiments \cite{ronnowQuantumPhaseTransition2005}; the lower mode is the electronuclear soft mode which governs the quantum critical behavior.

These modes are shown in Fig. \ref{fig:energy}(b), along with the modes determined by the Green's function. The two approaches are in good agreement, with the dominant spectral weight of the modes determined by the Green's function coinciding with the modes determined by the equation of motion. The primary discrepancy between the two approaches is in the vicinity of the quantum critical point, where the low energy mode softens to zero. This is to be expected because at the quantum critical point the electronuclear correlations dropped when the equation of motion is treated in the RPA will become important. In Sec. \ref{sec:domains}, we will use the equation of motion to calculate Walker modes present in the system; away from the quantum critical point the equation of motion is expected to give good results.


\section{Loop-Gap Resonators}
\label{sec:resonator}


As discussed above and as illustrated in Fig.~\ref{fig:energy}, the predicted electronuclear mode structure of \LiHoF\ has a level spacing of order 1-3 GHz (a few tens of $\mu$eV). Similarly, the dynamics of many magnetic materials, particularly collective modes in the close vicinity of a phase transition, are often found at energies below 0.1 meV. Standard techniques such as inelastic neutron scattering become increasingly difficult in this regime. Energies of order 10~$\mu$eV, corresponding to frequencies of a few GHz, are accessible via RF methods through the direct application of an ac magnetic field to the material. When studying quantum materials at milliKelvin temperatures and in dc magnetic fields, a number of experimental constraints and challenges arise. Standard resonant cavities tuned to frequencies of order $\sim$1--3~GHz have a characteristic size similar to their 10--30 cm wavelength, incompatible with the space constraints of typical superconducting solenoid magnets. Furthermore, achieving reasonable filling fractions and hence sensitivities in such cavities requires (preferably single-crystal) samples of similar size, which is impractical for many materials of interest.

An attractive solution to these constraints is the loop-gap resonator (LGR), developed from a magnetron topology for use in S-band EPR \cite{Froncisz:1982fw}, and subsequently applied to study a wide variety of physical systems including hybrid quantum devices \cite{ballLoopgapMicrowaveResonator2018, evertsMicrowaveOpticalPhoton2019}, molecular nanomagnets \cite{joshiAdjustableCouplingSitu2020}, diamond NV centers \cite{eisenachBroadbandLoopGap2018}, and even searches for axion dark-matter candidates \cite{mcallister3DLumpedLC2016}. As the name suggests, an LGR consists of a metal block in which a set of loops and gaps have been cut. The essential behavior of this construct is captured in a lumped-element circuit picture, where the loops are essentially single-turn inductors and the gaps are parallel-plate capacitors \cite{MEHDIZADEH:1983wa}. The electric field is largely confined to the gaps and the magnetic field to the loops. With the sample of interest placed in one of the loops, this results in a purely magnetic excitation. This separation of electric and magnetic fields is also important in minimizing dielectric heating of the sample, of special import at sub-Kelvin temperatures.

\begin{figure}
	\centering
	\includegraphics[width=2.75in]{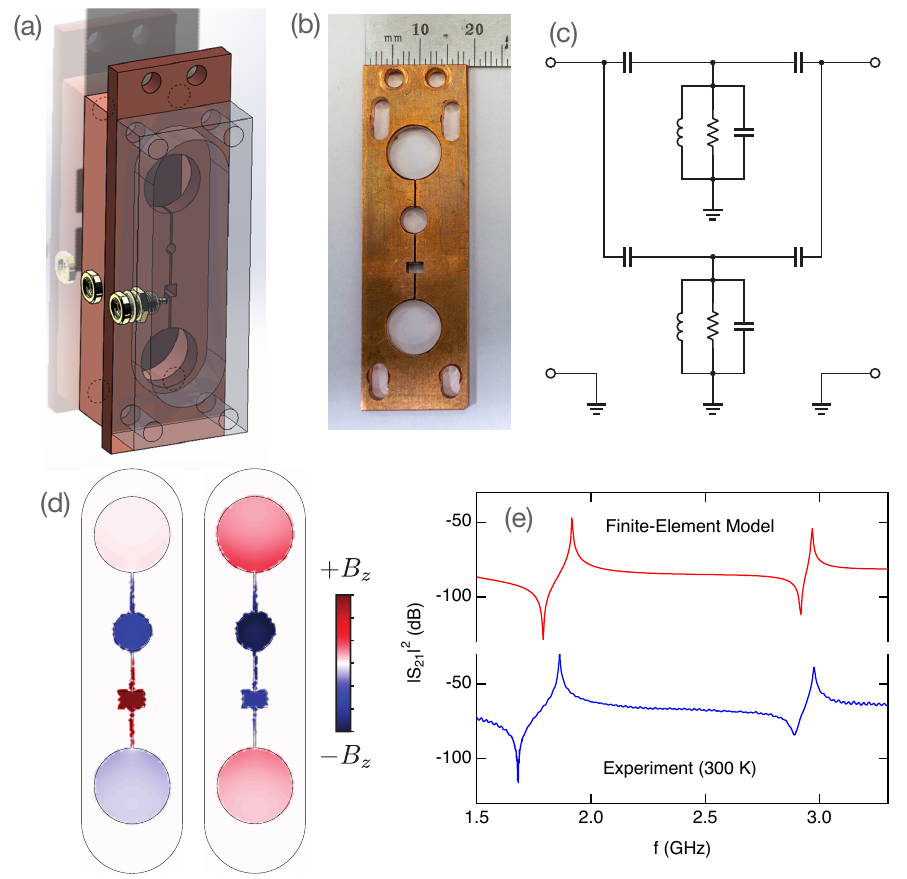}
	\caption{Resonator design and effective circuit (a) CAD rendering and (b) photograph of the 4-loop-3-gap (4L3G) resonator design. The contours are cut using wire electric discharge machining. The sample is placed in the rectangular loop, which has a very uniform AC magnetic field profile. The circular loops act as a return flux path, and the thin gaps ($\approx 330 \mu \mathrm{m}$) concentrate the electric field and provide capacitive tuning of the resonant modes by insertion of a sapphire wafer. Pins attached to two external MMCX coaxial connectors serve as input/output ports. The structure is made from oxygen-free high-conductivity (OFHC) copper, and has overall dimensions 60 x 20 x 3 mm. The resonator is enclosed in a copper shielding box to minimize radiation loss.  (c) Lumped-element circuit model of the 4L3G resonator. Each of the two parallel RLC circuits in the middle represent one of the modes, and the coupling is modeled with capacitors between the RLC circuits and input/output ports. (d) Finite-element calculation showing magnetic field density for the two primary modes of the resonator. (e) Calculated (top) and measured (bottom) frequency response for the multi-mode resonator, showing good agreement between the model and the physical device.}
	\label{fig:res}
  \end{figure}

While the simplest LGR design is a tube with a lengthwise slot providing the inductive and capacitive components respectively~\cite{Froncisz:1982fw}, the performance can be improved by employing a more complex geometry. For instance, a three-loop two-gap design offers advantages in suppressing radiation loss~\cite{Wood:1984dj}. Here, we use four-loop three-gap resonators (Fig.~\ref{fig:res}(a,b)) where the additional geometry yields two independent resonant modes at different frequencies. A lumped element circuit model for this system is shown in Fig.~\ref{fig:res}(c), with each mode represented by a parallel RLC circuit \cite{ballLoopgapMicrowaveResonator2018} and the input/output antenna ports are capacitively coupled to the resonator. To validate the lumped-element approach, the field patterns and frequency response of these resonators were calculated using the finite-element modelling package HFWorks~\cite{HFWorks2021}. The field patterns of the ac magnetic field of each of these dominant modes are plotted in Fig.~\ref{fig:res}(c). In the low frequency mode, the two central loops (of which the rectangular one holds the sample) are opposite in phase, with each central loop using  the larger outer loop as a return flux path. In the higher frequency mode, the two central loops are in phase, with the outer loops having the opposite phase of return flux, creating an overall quadrupole field pattern. A lower frequency ``dark'' mode which has negligible intensity at the sample location is not shown, nor are the higher frequency modes where the enclosure box halves form typical TE/TM resonant cavity modes. We demonstrate in Fig.~\ref{fig:res}(e) that the calculated frequency response is a good match to the actual measured behavior. Additional details on the finite-element modeling are given in Ref.~\cite{liberskyDesignLoopgapResonator2019a}.

A further feature of the LGR approach is its tunability. The active frequency can be varied over several GHz by introducing a dielectric material into the gaps, such as wafers of sapphire or alumina ceramic \cite{joshiAdjustableCouplingSitu2020}. This permits using one resonator to probe a range of frequencies. In the 4L3G configuration, the distribution of electric fields differs for each mode, allowing independent tuning of the modes \cite{liberskyDesignLoopgapResonator2019a}. The implementation of the 4L3G design used here has a total capacitance that can be varied over an order of magnitude: from 0.86 pF with no dielectric in the gaps to as high as 8.6 pF for the gaps completely filled with sapphire. The effective inductances are 1.1 and 0.3 nH for the lower and upper resonance modes, respectively.   For situations where multiple modes introduce unwanted complexity, the simpler 3L2G geometry may be preferable. We used such a single-mode cavity with maximum capacitance of 51 pF and 0.5 nH inductance at our lowest probe frequencies ($<1$~GHz).

With a sample in the resonator, any non-zero magnetic susceptibility will change the effective inductance and hence the resonant frequency and quality factor of the overall circuit. Here, we derive a general expression for the response of the circuit; in the following sections we combine this general form with the specifics of the \LiHoF\ Hamiltonian to obtain a model that can be compared directly to the experimental data. Near resonance ($\omega \approx \omega_r = 1/\sqrt{LC}$), one may write the complex impedance of an individual parallel RLC circuit as \cite{pozarMicrowaveEngineering2012}
\begin{align}
Z_{RLC} = \frac{R}{1+2j Q (\omega/\omega_r-1)},
\end{align}
where the quality factor of the resonator is $Q=\omega_r/\gamma$, with
$\gamma=1/(RC)$. For a sample with complex susceptibility $\chi (B,\omega)$ and
filling factor $\eta$, the inductance becomes $L \rightarrow L' = \left[1 +\eta^2
\chi(B, \omega)\right]L$. Assuming that $|\eta^2 \chi| \ll 1$,
\begin{align}
\label{eq:ZRLC}
Z_{RLC} = \frac{R\gamma/2e^{-j\pi/2}}{\omega-(\omega_r-g^2 \chi')-j(\gamma/2 + g^2
\chi'')},
\end{align}
where the susceptibility has been separated into its real and imaginary parts, $\chi=\chi'-j\chi''$ (the minus sign is conventional in the electrical engineering literature). In terms of the circuit parameters, the coupling strength is $g^2=\eta^2 \omega_r/2$. As one approaches the system's critical point, the susceptibility diverges, and the condition $|\eta^2 \chi| \ll 1$ no longer holds. The enhanced inductance of the RLC circuit drives the resonator mode to zero, $\omega_r \rightarrow 0$ as $L' \rightarrow \infty$, at which point one expects the light-matter system to undergo a superradiant quantum phase transition. However, this neglects the effects of damping and decoherence of the LiHoF$_4$ system which may prevent superradiance \cite{mckenzieTheoryMagnonPolaritons2022}.

In a two-port microwave network, a resonator mode can be modeled as a T-network, with the antenna impedances $Z_a$ on the series branches, the transmission lines at the input and output ports each having impedance $Z_0$, and the parallel resonator impedance being $Z_{RLC}$. Deriving the S-parameters for such a circuit is straightforward \cite{pozarMicrowaveEngineering2012}. The result for the
transmission function is
\begin{align}
\label{eq:ResTransCircuit}
S_{21}&=\frac{Z_0/(Z_a+Z_0)}{1+(Z_a+Z_0)/(2Z_{RLC})} \\ \nonumber
&\approx \frac{Ae^{-j\theta}}{\omega-(\omega_r-g^2 \chi')-j(\Gamma_r/2 + g^2 \chi'')}.
\end{align}
In the final expression, we have made use of Eq. (\ref{eq:ZRLC}). The overall amplitude and phase of the transmission function may depend on attenuation in the transmission lines, the reference planes of the input and output ports, and other details of the microwave resonator \cite{pozarMicrowaveEngineering2012, Probst}; here, we leave $A$ and $\theta$ as unspecified parameters. The damping of the resonator mode is related to the damping of the RLC circuit by $\Gamma_r = \gamma[1+2R/(Z_a+Z_0)]$. In Section \ref{sec:MPhybridization}, we compare the transmission function obtained from the circuit model with the transmission function obtained from a microscopic treatment of the magnon-photon system \cite{harderStudyCavitymagnonpolaritonTransmission2016,mckenzieTheoryMagnonPolaritons2022}.

\begin{figure}
\includegraphics[width=3.4in]{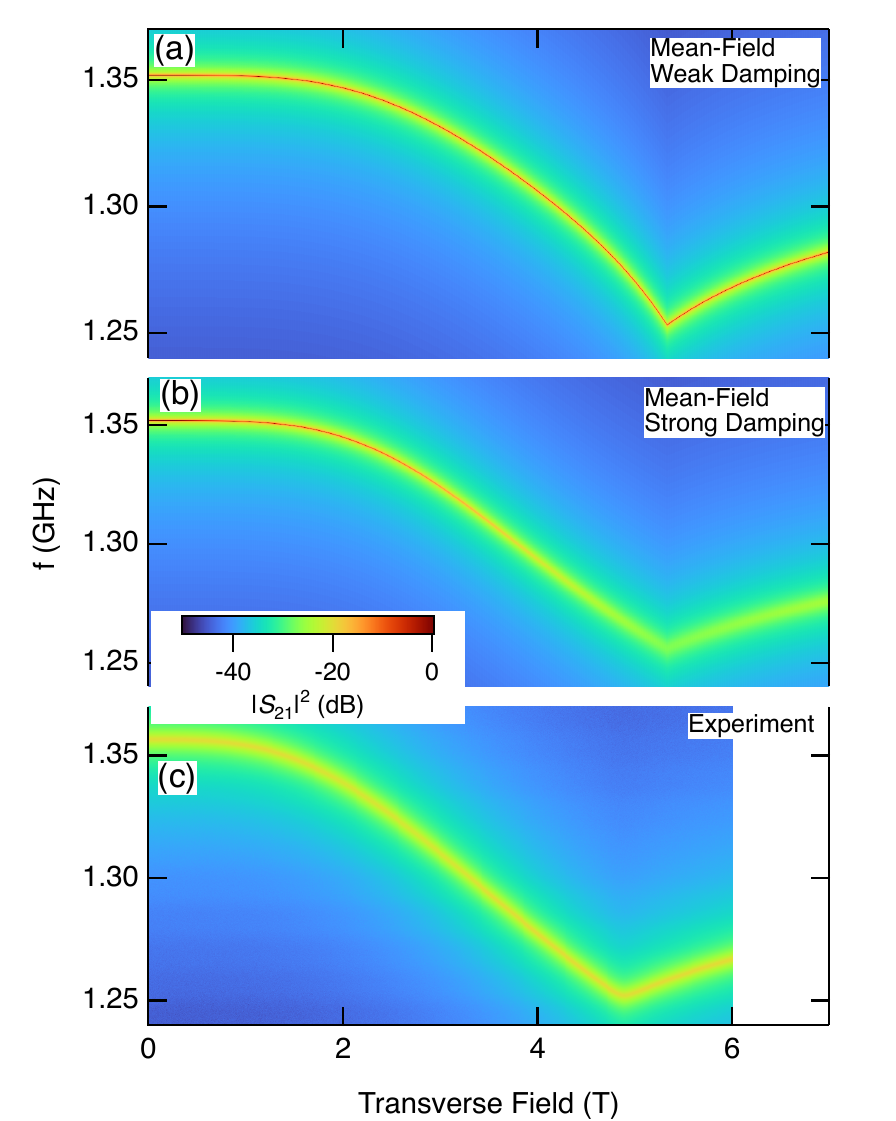}
\caption{Mean-field susceptibility calculations and the effects of damping. (a) Transmission coefficient $|S_{21}|^2$, using a mean-field solution to (\ref{eq:LiHoTrunc}) and the circuit-model expression for the S-parameters, (\ref{eq:ResTransCircuit}) in the weak-damping limit with $\Gamma_r=1\:\mu\mathrm{K}$ and $\eta=0.67$. (b) $|S_{21}|^2$ calculated in the strong-damping limit, $\Gamma=500\:\mathrm{mK}$, $\eta=0.8$. (c) Experimental $|S_{21}|^2$ at $T=55$~mK. The evolution of the mode, and in particular the essentially linear approach to the quantum phase transition at $H_t=4.9$~T, are best matched by the calculations shown in (b), suggesting that the system is in a strong-damping limit.}
	\label{fig:lowfreq}
\end{figure}

To connect these calculations with measurements, we plot in Fig.~\ref{fig:lowfreq} the transmitivity $|S_{21}|^2$, inserting into (\ref{eq:ResTransCircuit}) mean-field solutions to (\ref{eq:LiHoTrunc}). We examine  weak and strong damping of the lowest MF mode, $\Gamma_1$ of $1\:\mu\mathrm{K}$ and 500~mK (panels a and b respectively), and compare with the measured $S_{21}$ (panel c). The filling factors $\eta$ in panels a and b (0.67 and 0.8, respectively) were tuned to match the overall span of the data, and the damping of all other modes was set to $1\:\mu\mathrm{K}$.  In comparing the mean-field calculations to the measurements, we note first that MF solutions are known to overestimate the critical field for the quantum phase transition in \LiHoF\ \cite{ronnowMagneticExcitationsQuantum2007}. We examine instead the shape of the resonator response on the approach to the phase transition. As shown in Fig.~\ref{fig:lowfreq}(c), the peak resonance frequency evolves approximately linearly as a function of transverse magnetic field. Comparing with the calculations, it is clear that the strong-damping limit shown in (b) best reproduces the observed behavior. More in-depth comparisons between measurement and calculation, including the full RPA description from above, are presented below.

We note also that by measuring the response with a vector network analyzer, one may probe both the amplitude and phase of the transmission function.  No new information is present in the phase data; however, in some instances it provides better resolution of the resonator modes. In a bimodal resonator, assuming that the resonator modes are weakly coupled, the total transmission is found by summing the transmission functions of the two individual resonators, $S_{21} = S_{21}^a + S_{21}^b$, where, in the absence of the magnetic material,
\begin{align}
S_{21}^{a,b} = \frac{A_{a,b} e^{-j \theta_{a,b}}}{\omega - \omega_{a,b}
-j\Gamma_{a,b}/2}.
\end{align}
Assuming the phase shift of the two modes is the same, $\theta_a = \theta_b$, one finds a minimum (an anti-resonance) in the total transmission function at
\begin{align}
\omega_{anti} = \frac{A_a \omega_b + A_b \omega_a}{A_a + A_b}.
\end{align}
If the amplitudes of the two modes are the same, the anti-resonance occurs at their average value. One may tune the location of the anti-resonance by adjusting the amplitudes of the modes, as illustrated in Section~\ref{sec:Spectra} below.


\section{Magnon-photon hybridization}
\label{sec:MPhybridization}


In the presence of strong coupling between light and matter, one must consider hybridization of the modes. In the context of microwave cavity photons and spin excitations, these hybridized excitations are known as cavity magnon-polaritons \cite{Rameshti}. Commonly studied in the context of yttrium iron garnet (YIG) crystals, these systems can display a variety of interesting phenomena including magnetically induced transparency \cite{wangMagnoninducedTransparencyAmplification2018}.

The behavior of coupled magnon-photon modes is characterized by the cooperativity parameter $C = g_m^2/\kappa \gamma$ \cite{hueblHighCooperativityCoupled2013}, which indicates the coupling strength relative to dissipation sources. For $C>1$ we enter the strong coupling regime where mode splitting becomes prominent, resulting in avoided level crossings. At the experimentally relevant temperatures and frequencies, we may drop the elastic contribution to the dynamic susceptibility discussed following equation (\ref{eq:GF}), and write down an effective bosonic Hamiltonian describing the magnon-photon interactions \cite{mckenzieTheoryMagnonPolaritons2022}:
\begin{align}
\label{eq:Hmp}
\mathcal{H}_{mp} = \omega_r &a^{\dagger} a  + \sum_{m=1}^M \omega_m b_m^{\dagger} b_m \\ \nonumber
&+ (a^{\dagger}+a) \sum_{m=1}^M g_m (b_m^{\dagger}+b_m).
\end{align}
In terms of the spin-photon coupling strength $\alpha = \eta \sqrt{2\pi} \sqrt{\hbar \omega_r} \sqrt{\rho_s J_D}$, the magnon-photon coupling strength is given by $g_m^2 = \alpha^2 A_m$, where $A_m$ is the spectral weight of the magnon mode. One may diagonalize this Hamiltonian to obtain the magnon-polariton modes of the system $\mathcal{H}_{mp} = \sum_{m=1}^{M+1} \Omega_m d_m^{\dagger} d_m$. At finite temperatures, considering the truncated \lhf\ Hamiltonian, one has $M=120$ with magnon modes corresponding to all the possible transitions between the $16$ low energy electronuclear modes. Many of these modes will be suppressed thermally, or will carry negligible spectral weight, and may be neglected in the calculations.

From a microscopic description of the spin-photon system \cite{mckenzieTheoryMagnonPolaritons2022}, one may calculate the imaginary time magnon-polariton propagator, defined by $D_{mp}(\tau) = \bigr\langle T_{\tau} \bigr(a^{\dagger}(\tau)+a(\tau)\bigr) \bigr(a^{\dagger}+a\bigr) \bigr\rangle$. At Matsubara frequency $z=i\omega_n=2\pi i n/\beta$, the propagator is given by
\begin{align}
\label{eq:MPprop2}
D_{mp}(z) = -\frac{2\omega_r}{\beta} \biggr[\frac{1}{z^2 - \omega_r^2
+ \alpha^2 2\omega_r \chi_J(z)}\biggr].
\end{align}
We express $D_{mp}(z)$ in terms of the dynamic spin susceptibility [see Eq. (\ref{eq:ChiSpin})] rather than the susceptibility of the material used in the circuit model analysis. This propagator is related to the response function of the cavity photons by $D_{mp}^{ret}(\omega) = \beta D_{mp}(z \rightarrow \omega + i0^+)$. The imaginary component of the response function determines the energy absorbed by the resonator photons, which we assume to be proportional to the resonator transmission function $|S_{21}|^2 \propto \text{Im}[D_{mp}^{ret}(\omega)]$ \cite{harderStudyCavitymagnonpolaritonTransmission2016}. Using the spectral representation of $\chi_J(z)$ [Eq. (\ref{eq:ChiSpin})] and comparing Eq. (\ref{eq:MPprop2}) with the propagator obtained from Eq. (\ref{eq:Hmp}), one obtains the result $g_m = \alpha^2 A_m$.

One may include phenomenological damping of the resonator mode, in which case the resonator transmission is given by
\begin{align}
\label{eq:TransMicro}
|S_{21}|^2 \propto \text{Im}[D_{mp}^{ret}] =\frac{2\omega\omega_r\Gamma_{mp}}{(\omega^2-\omega_{mp}^2)^2 +
(\omega\Gamma_{mp})^2},
\end{align}
where
\begin{align}
\omega_{mp}^2 = \omega_r^2 + (\Gamma_r/2)^2 - 2 \alpha^2 \omega_r \chi_J'(\omega)
\end{align}
determines the magnon-polariton modes of the system, and the damping function is
\begin{align}
\omega \Gamma_{mp} = \omega \Gamma_r + 2 \alpha^2 \omega_r \chi_J''(\omega).
\end{align}
The factor of $(\Gamma_r/2)^2$ in the mode equation is a counter-term which eliminates a shift in the resonator frequency due to its damping.

Recall that from the circuit model the transmission function [Eq. (\ref{eq:ResTransCircuit})] is given by
\begin{align}
\label{eq:S21circuit}
|S_{21}|^2 \propto
\frac{1}{[\omega-(\omega_r-g^2 \chi'(\omega)]^2 + [\Gamma_r/2+g^2
\chi''(\omega)]},
\end{align}
where $\chi(\omega) = 4\pi \rho_s J_D \chi_J(\omega)$. Comparing Eq. (\ref{eq:S21circuit}) to Eq. (\ref{eq:MPprop2}), one sees $g^2 = \alpha^2 2\omega_r/(4\pi \rho_s J_D) = \eta^2 \omega_r/2$, so the coupling in the microscopic theory is in agreement with the coupling determined by the circuit model. The transmission function given in Eq. (\ref{eq:TransMicro}) differs from the transmission function determined by the circuit model in two important ways: 1) The transmission function determined by the resonator photon response function contains counter rotating terms not present in the circuit model. These terms may be important when the magnon and photon modes are strongly coupled. 2) In the circuit model, at the magnon mode energies where $\chi''$ becomes large, the resonator transmission is attenuated. In Eq. (\ref{eq:TransMicro}), the factor of $\Gamma_{mp}$ in the numerator contains a term proportional to $\chi''$, so there will be resonant transmission of photons at the magnon mode energies.

\begin{figure}[tb]
\centering
\includegraphics[width=3in]{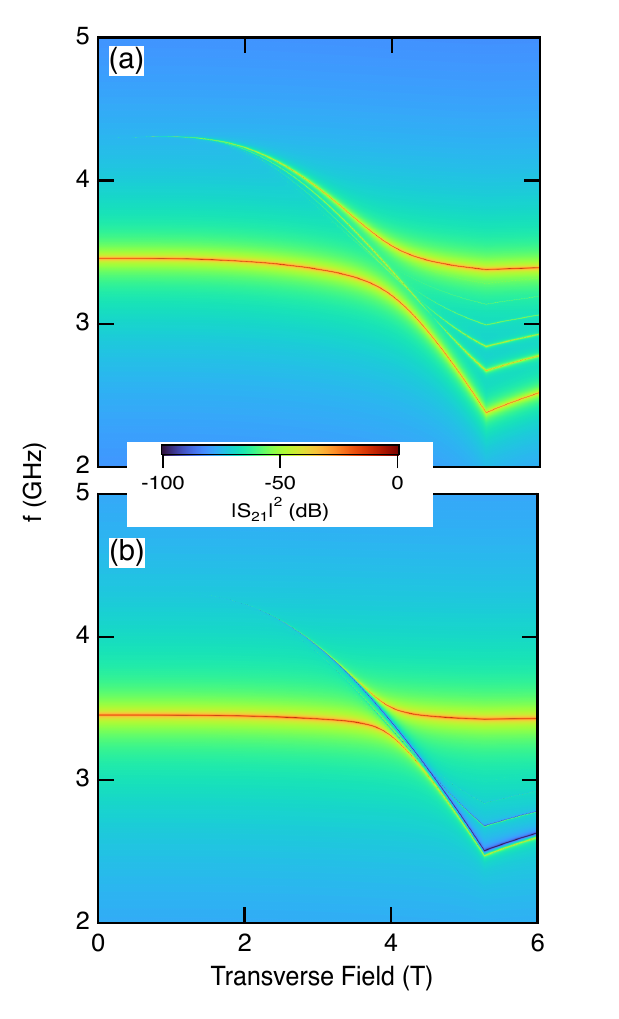}
\caption{Transmittance of the combined resonator-\LiHoF\ system calculated using (a) the microscopic model (\ref{eq:TransMicro}), and (b) the circuit model (\ref{eq:ResTransCircuit}). The calculations are done at $T=50$ mK, with the filling factor set to $\eta=0.25$. The LiHoF$_4$ susceptibility is calculated in MF theory. Dampings of the resonator mode and the magnon modes are set to a fixed constant $\Gamma=1\mu \mathrm{K} \approx 20.8$ MHz.}
\label{fig:TransComp}
\end{figure}

We compare in Fig. \ref{fig:TransComp} the single-ion (MF) resonator transmission function calculated from the circuit model and from the microscopic theory which includes counter-rotating terms. Transmission by the magnon modes is stronger in the microscopic theory due to the factor of $\chi''(\omega)$ in the numerator of the transmission function. The frequency dependence of the numerator leads to a larger avoided level crossing in the microscopic transmission data than in the circuit model, and the minimum of the lower polariton mode in the circuit model is at $f_\mathrm{min} \approx 2.57$~GHz, whereas in the microscopic model $f_\mathrm{min} \approx 2.44$~GHz. In an experiment, one expects some of the energy absorbed by the damped magnon modes to contribute to resonator transmission, and some of the energy to be lost to the system's environment; a realistic model of the resonator transmission lies somewhere in between these two models. In what follows, we will make use of Eq. (\ref{eq:TransMicro}) to model the resonator transmission as this function more accurately reflects what is seen in the experiments.

The MF and RPA models have comparable results for $f>3.5$~GHz, whereas a significant discrepancy is seen at lower frequencies where the lowest energy magnon mode is seen to soften to zero in the RPA. The divergent spectral weight of the soft mode leads to strong coupling, and, in the absence of dissipation and decoherence, drives the lower polariton mode to zero, at which point the system undergoes a superradiant phase transition. In experiment, dissipation and decoherence forestall this transition, and the soft mode shows up as a weak avoided level crossing, or a resonance in the dissipation of the lower polariton mode \cite{mckenzieTheoryMagnonPolaritons2022}. This is discussed explicitly in the following section.


\section{The Role of Soft Photons}
\label{sec:SoftPhotons}


A naive treatment of the transmitivity in our experiment predicts that it will diverge at the resonance associated with the soft mode, when the frequency of the soft mode (which couples to photons) goes to zero. This is incorrect, for reasons we have introduced in our previous work \cite{LiberskySM,mckenzieTheoryMagnonPolaritons2022}. This divergence is reminiscent of what happens in any calculation of the effect of soft photons on electronic processes, a problem much discussed over the years.

Here, we give some more general insight into the physics. We start by recalling key features of infrared (IR) divergences in Quantum Electrodynamics (QED), and then see what this implies for photons in a cavity.

\subsection{Soft Photons in Vacuum QED}

Naive calculations in QED immediately give IR  divergent cross-sections, with an infinite number of emitted soft photons \cite{blochNoteRadiationField1937,peskinSections1995}. This is an apparent failure of perturbation theory; the IR limit describes classical processes like Bremsstrahlung, in which an electron scatters off some potential, and which have a well-behaved low energy cross-section. Thus, one immediately has a contradiction between naive theory and the real world.

\subsubsection{The ``Soft Factor''}

\begin{figure}
    \centering
    \includegraphics[width=3.5in]{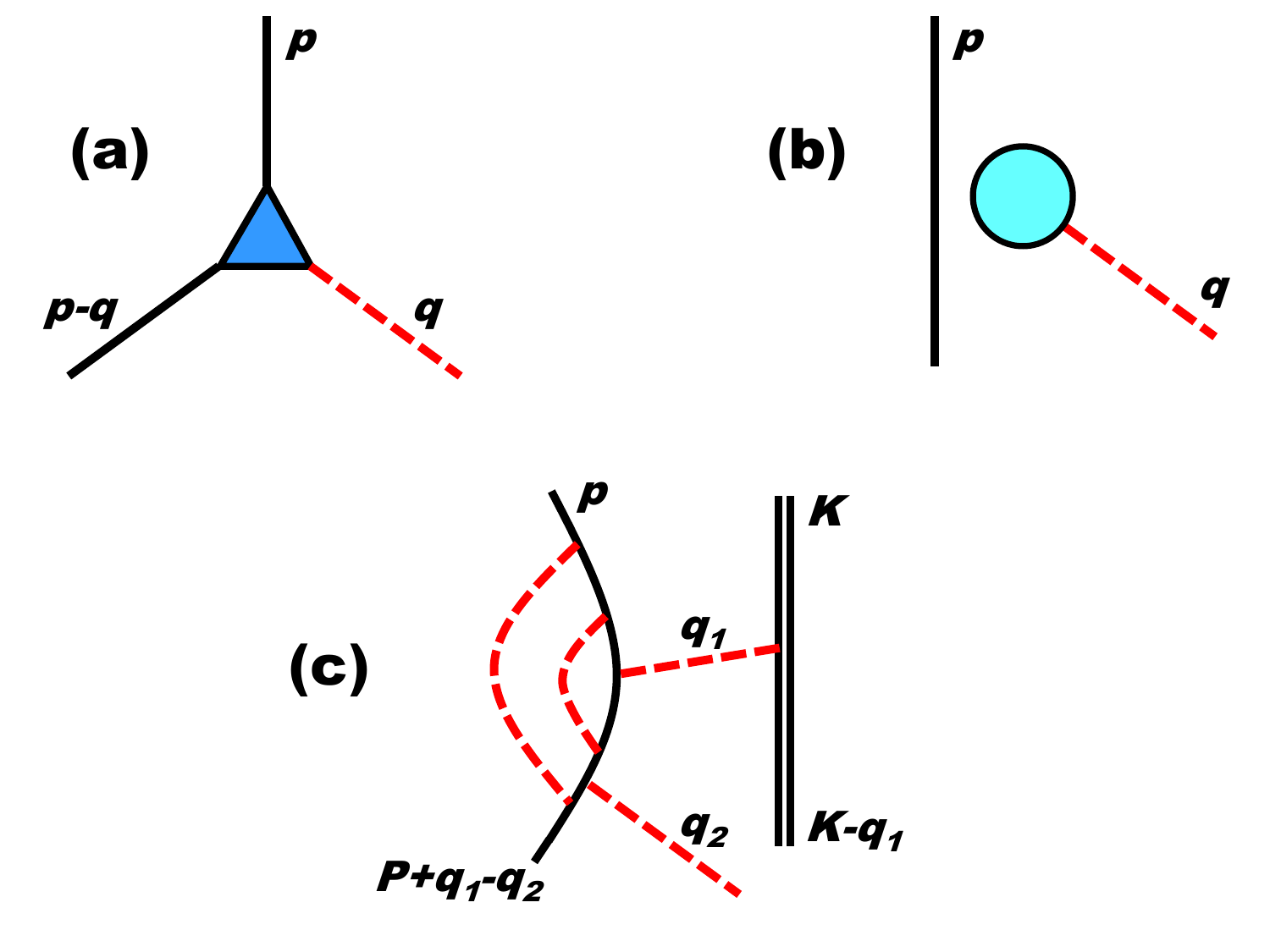}
    \caption{In (a) we show the 3-point vertex $\Lambda(p,q)$ for electron-photon scattering, which in (b) is shown in the low-energy IR regime, where it factorizes as given in equations (\ref{eq:Lambda}) and (\ref{eq:Sq}). In (c), we depict a contribution to the scattering of an electron by an external potential, in which vertex corrections to photon emission occur; these cancel
    the divergences shown in (a).}
    \label{fig:photon}
\end{figure}

The paradox can be resolved in various ways. Consider, as a simple example, the scattering process in Fig.~\ref{fig:photon}(a), where an electron with 4-momentum $p$ emits a soft photon with 4-momentum $q$. Then the ``factorization theorem''\cite{sudakovVertexPartsVery1956,abrikosovInfraredCatastropheQuantum1956} states that in the IR limit, as $q\rightarrow0$, this process is described by a scattering amplitude $\Lambda_3(p,q)$. To leading order, this has the form
\begin{equation}\label{eq:Lambda}
    \Lambda_3(p,q)\rightarrow K_2(p)S(q),
\end{equation}
in which $K_2(p)$ is the usual electron propagator (fully renormalized), and where $S(q)$ is the ``soft factor'', having the behavior
\begin{equation}\label{eq:Sq}
    S(q)\sim S_0(q)+S_1(q)+O(|q|).
\end{equation}

In this equation, $S_0(q)\sim O(1/|q|)$ is the ``leading term'' (in the sense of an asymptotic series), and it is divergent in the $q \rightarrow 0$ limit. The next term $S_1(q)\sim O(1)$ is the ``sub-leading term''; after this one has ``sub-sub-leading terms,'' which we do not discuss here.

There is a divergence as $q \rightarrow 0$, and the question is how to cure it. The divergence arises from the infinite number of soft photons that are emitted when the electron is subject to any acceleration (e.g., from an interaction with some heavy body which acts as an external potential). If, however, we wish to calculate the scattering cross-section of the electron off this body, we must include vertex corrections to the scattering of the electron, as shown in Fig.~\ref{fig:photon}(c), if we are to get a consistent result. One then finds a cancellation in the cross-section for scattering, to leading order, between the divergences associated with photon emission in $S_0(q)$, and the divergences in the vertex corrections \cite{kinoshitaNoteInfraredCatastrophe1950,nakanishiGeneralTheoryInfrared1958}. In this way one can recover classical Brehmsstrahlung.

One also can recover these results, as well as the form for $\Lambda_3(p,q)$, using an eikonal treatment,\cite{fradkinApplicationFunctionalMethods1963,fradkinApplicationFunctionalMethods1966} which is more elegant in that it is already a non-perturbative approach. More modern treatments of these IR phenomena begin from fully dressed coherent states for the electron \cite{kibble1969,kulish1970,kapecInfraredDivergencesQED2017}. Physically this makes sense because one can never undress the photons from the electron. For an electron in an infinite flat vacuum, one also can relate the form of $S(q)$ to asymptotic symmetries of QED \cite{campigliaAsymptoticSymmetriesQED2015, choiBMSSupertranslationSymmetry2018}.

Moving from the leading divergent term to $\Lambda_3(p,q)$ (or any other process involving electrons and soft photons) to the contribution of sub-leading terms \cite{lysovLowSubleadingSoft2014,campigliaSubleadingSoftPhotons2016,choiSubleadingSoftDressings2019,delisleSoftTheoremsBoundary2021} introduces complications. These sub-leading terms have been analyzed in various ways: as a manifestation of ``large'' gauge transformations\cite{campigliaSubleadingSoftPhotons2016} or of the boundary terms in a path integral for the scattering amplitude \cite{delisleSoftTheoremsBoundary2021,delisleDecoherence2PathSystem2022}. We emphasize here that, even now, there is no general agreement as how the sub-leading terms should be understood, and how they may enter into physical results; we return to this point below.

\subsubsection{The Influence Functional}

Another perspective, which is more directly connected to experiment, is to look directly at the dynamics of the reduced density matrix $\rho(x,x^\prime)$ for the electron \cite{feynmanTheoryGeneralQuantum2000}. In abbreviated notation \cite{wilson-gerowFunctionalApproachSoft2018}, where, e.g.,  we write $\rho(1,1^\prime) \equiv \langle 1| \rho |1^\prime \rangle$, the equation of motion for this density matrix $\rho$ can be written in the form
\begin{equation}\label{eq:density_matrix}
    \rho(2,2^\prime) = \int d1\:d1^\prime \; K(2,2^\prime;1,1^\prime) \;
    \rho(1,1^\prime),
\end{equation}
where the propagator $K(2,2^\prime;1,1^\prime)$ for $\rho$ has the path integral form\cite{wilson-gerowFunctionalApproachSoft2018}
\begin{equation}
 \label{eq:propagator}
    K(2,2^\prime;1,1^\prime) = \int_{j_1}^{j_2}\mathcal{D}j
    \int_{j^\prime_1}^{j^\prime_2}\mathcal{D}{j^\prime}
    e^{i(S_0[j]-S_0[j^\prime])/\hbar}\; \mathcal{F}[j,j^\prime],
\end{equation}
in which $S_0[j]$ is the bare action for the electron (i.e., without the coupling to the EM photon field), written in terms of the electron 4-current $j^\mu(x)$, and where $\mathcal{F}[j,j^\prime]$ is the influence functional, describing the way in in which the functional integration over photons weights different 4-current paths $\{ j, j^{\prime} \}$ for the density matrix propagator.

For QED in a vacuum at finite temperature $T$, this influence functional has the form
\begin{equation}
    \mathcal{F}[j,j^\prime] = e^{i(\Delta[j,j^\prime]+i\Gamma[j,j^\prime])/\hbar},
\end{equation}
where the imaginary part of the phase of the influence functional, known as the ``decoherence functional'', is of particular interest here. It has the form
\cite{breuerDestructionQuantumCoherence2001}
\begin{equation}
    \Gamma[j,j^\prime] \;=\; \frac{e^2}{2}\int \frac{d^3
    q}{(2\pi)^3}\frac{1}{2\omega} \;\mathcal{P}^{\mu\nu}(q)\,
    \delta j_\mu(q)\delta j_\nu(q)\, \coth\frac{\hbar\beta\omega}{2},
\end{equation}
where $\delta j_\mu = j_\mu(q)-j_\mu^\prime(q)$ and $\mathcal{P}_{\mu\nu}=\delta_{ij}-q_iq_j/|q|^2$ (for $i,j=1,2,3$) and $\mathcal{P}_{0\nu}(q)= \mathcal{P}_{\mu0}(q)=0$.

In studying the contribution of soft photons to the decoherence functional one begins by separating $j_\mu(q)$ into ``hard'' (large $q$) and ``soft'' (small $q$) parts. At the level of the leading divergent terms, the correct way to dress the $j_{\mathrm{soft}}^\mu (q)$ is well understood, and is summarized in the soft factor $S_0(q)$ in equations \ref{eq:Lambda} and \ref{eq:Sq}.

However, just as found when analyzing the soft factor, the  way to handle the sub-leading term $S_1(q)$ is less clear. Elsewhere, we have argued \cite{delisleDecoherence2PathSystem2022} that it is not necessary to dress $j^{\mathrm{soft}}_\mu (q)$ with sub-leading terms.  However, it would appear that the only way to determine whether it should be dressed or not is by doing experiments - there is as yet no obvious theoretical criterion we can use to decide.

In principle one can do experiments on electronic decoherence rates. In Ref. \cite{delisleDecoherence2PathSystem2022} we discuss two such experiments; it is noteworthy that in both cases (a 2-slit experiment, and an optomechanical experiment), one requires a theory of the interaction of the electron with the surrounding solid media. Thus, any theory of the sub-leading contributions to electronic decoherence already requires that we go beyond vacuum QED, and consider objects like cavities (each slit on the 2-slit experiment constitutes a kind of cavity for the electron \cite{delisleDecoherence2PathSystem2022}). It seems that in the discussion of the IR properties of vacuum QED, we are inevitably led to consider cavity QED.

\subsection{Soft Photons in a Cavity}

In general we are dealing here with an optical cavity that supports both a driving laser mode and internal degrees of freedom -- the ones of interest here being a combination of photons and electronuclear magnons, already discussed above. This problem is similar to the standard optomechanical system, in which the laser drives both internal optical modes and mechanical (mirror) modes \cite{aspelmeyerCavityOptomechanics2014}, and where a set of oscillators represents the photon modes (this can be supplemented, if necessary, by the introduction of a set of ``spin bath'' modes \cite{Prokofev:2000bd}, describing localized environmental modes in the cavity, such as two-level systems or spin impurities).

In the present case, we have a cavity with an interaction $\mathcal{H}_{mp} $ of strength $g_m$ between the $m$-th electronuclear magnon mode and the photons (cf. eq. (\ref{eq:Hmp}) above); we recall that $g_m^2 = \alpha^2 A_m$, where $A_{m} = \lim_{k\rightarrow0} A_m(k)$ is the $k\rightarrow0$ limit of the spectral weight for the m-th electronuclear mode and $\alpha$ is the spin-photon coupling.

There are three things that we should like to emphasize here:

(i) As already noted in our previous work \cite{LiberskySM}, our system can be mapped to a standard Caldeira-Leggett model \cite{caldeiraQuantumTunnellingDissipative1983}. With the introduction of a suitable coordinate $Q$ for the magnon mode of interest (here the zero mode), one simply has a problem of an oscillator coupled to an oscillator bath \cite{LiberskySM,mckenzieTheoryMagnonPolaritons2022}, for which the decoherence functional has the usual form
\begin{multline}
\label{eq:decoherence_functional}
    \Gamma(Q,Q^\prime) = \int_0^\infty \frac{d\omega}{\pi\hbar} J(\omega)\int^t
    d\tau_1 \int^{\tau_1} d\tau_2
    \cos\omega(\tau_1-\tau_2)\\ \, \coth\frac{\beta\hbar\omega}{2}\;
    (Q(\tau_1)-Q^\prime(\tau_1) \, (Q(\tau_2)-Q^\prime(\tau_2)),
\end{multline}
where the ``Caldeira-Leggett spectral function'' $J(\omega)$ in principle can be extracted directly from experiment. The oscillators in the oscillator bath, with frequencies $\{ \omega_q \}$, represent the photons which damp the electronuclear zero mode.

(ii) The Caldeira-Leggett spectral function $J(\omega)$ and the zero mode damping $\gamma_0$ can be written as \cite{LiberskySM}
\begin{equation}
    J(\omega)=\frac{\pi}{2}\sum_q {g^2 \over \omega_q}\delta(\omega-\omega_q)
\end{equation}
and
\begin{equation}
    \gamma_0 (\omega)=\lim_{\epsilon\rightarrow0^+}\sum_q {g^2 \over \omega_q}
    \frac{i \omega^2}{\omega^2-\omega_q^2 + i\omega \epsilon}.
\end{equation}
Both the coupling $g^2$ to the zero mode and the zero mode damping $\gamma_0$ diverge as we approach the critical field, to produce the finite result for $S_{21}(\omega)$ that is seen in the experiment.

(iii) The dynamic susceptibility $\chi(\omega)$ is written as
\begin{equation}\label{eq:chi_theory}
    \chi(\omega) = \sum_m A_m \left[
        \frac{1}{\omega + E_m + i\gamma_m/2} - \frac{1}{\omega - E_m +
        i\gamma_m/2},
    \right]
\end{equation}
where $\gamma_m$ is the damping coefficient for the $m$-th electronuclear magnon mode.

We can now ask what, if any, is the relationship between the divergences enumerated here, in $g_0$ and in $\gamma_0$, and the divergences encountered in QED in the IR limit? The answer to this question is not obvious because the diagrammatic approximation schemes used in strongly correlated condensed matter physics are not typically designed to extract the leading terms in an asymptotic expansion about zero energy. Instead, they are set up so that all conservation laws (Ward identities) for the system are obeyed. In practice one employs a ``conserving approximation'' \cite{baymConservationLawsCorrelation1961,baymSelfConsistentApproximationsManyBody1962} relating the electron propagator $K_2(p)$, the 3-point vertex $\Lambda_3(p,q)$, the 4-point electron scattering function $\Gamma(p,p';q)$, and the irreducible 4-point vertex ${\cal I}(p,p')$ (noting that such conserving approximations still violate crossing symmetry \cite{stampSpinFluctuationTheory1985}).

There are notable exceptions to this: for example, analyses have been done for the Kondo problem using either a mapping to the spin-boson system, with $J(\omega)$ taking Ohmic form \cite{leggettDynamicsDissipativeTwostate1987a}, or written in such a way that one sums infinite classes of diagrams in which one sees the cancellations between vertex and self-energy terms \cite{verwoerdFeynmangraphTheoryKondo1974,verwoerdFeynmangraphTheoryKondo1974a}. In the same way, one can derive a form for $J(\omega)$ for QED consistent with the leading contributions to low-energy physical processes \cite{breuerDestructionQuantumCoherence2001}. However, in a Caldeira-Leggett formulation the complexity of the cancellation between vertex corrections and radiative corrections is actually concealed  -- we only see the final result.

\begin{figure}
\centering
\includegraphics[scale=0.35]{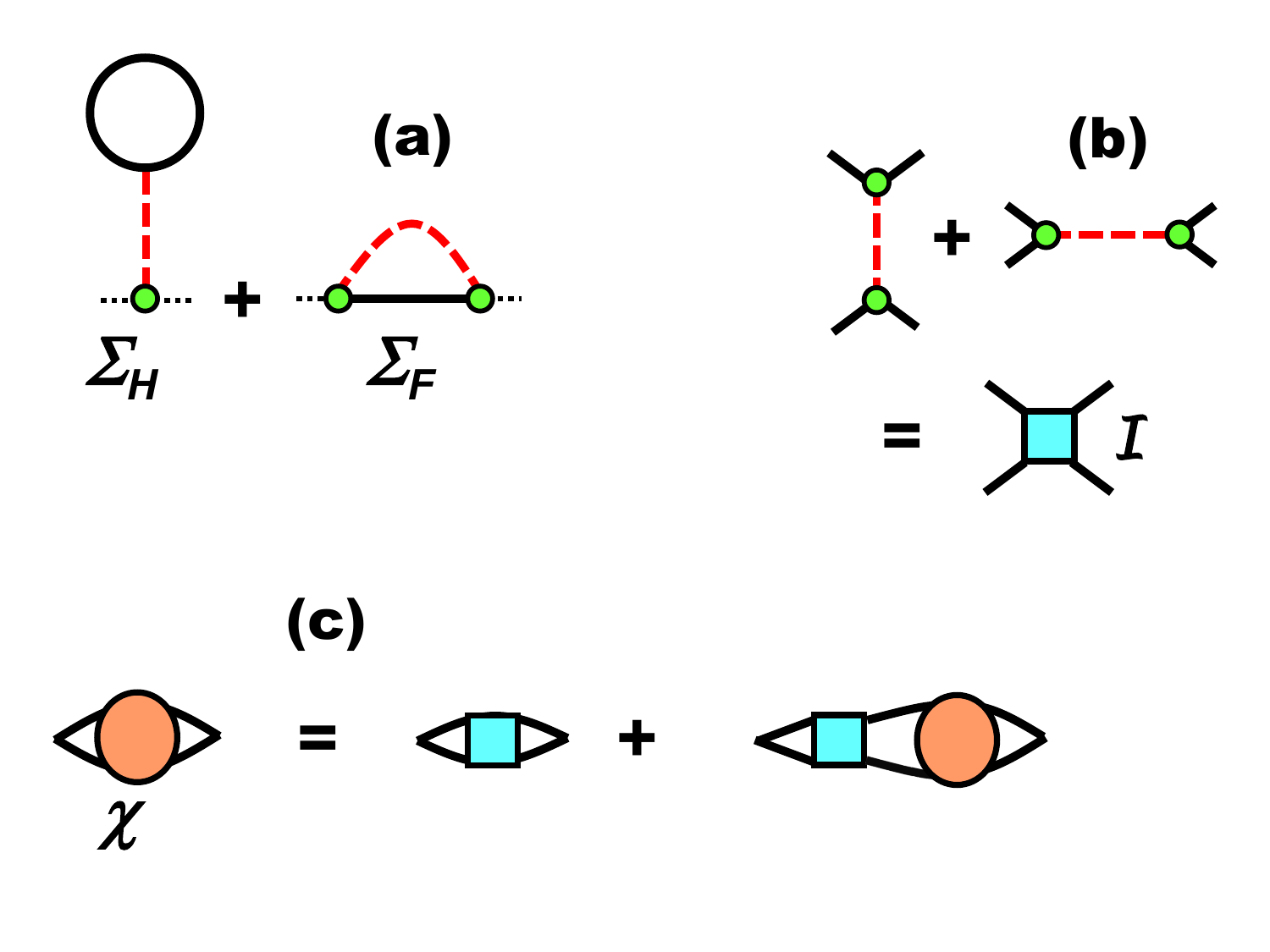}
\vspace{-7mm}
\caption{ The Hartree-Fock/RPA approximation used to calculate the dynamic susceptibility $\chi(\omega)$ in the text. (a) The self-energy. The solid line is the HF propagator $G_o$, and the hatched line is the renormalized photon propagator ${\cal D}$. (b) The resulting irreducible 4-point vertex ${\cal I}$, containing both Hartree and Fock terms. (c) The resulting Dyson equation for the dynamic susceptibility $\chi(\omega)$.           }
 \label{fig:RPA}
\end{figure}

The Random Phase Approximation (RPA) which we have used to describe the electronuclear collective modes in the present system is actually a very well-known and simple conserving approximation. One assumes a Hartree-Fock form for the self-energy $\Sigma ({\bf p}, \epsilon)$ (see Fig. \ref{fig:RPA}(a)), which leads to the irreducible 4-point vertex ${\cal I}({\bf p}, {\bf p'}, \epsilon, \epsilon')$ shown in Fig. \ref{fig:RPA}(b), and the dynamic susceptibility $\chi({\bf q}, \omega)$ shown in Fig. \ref{fig:RPA}(c) (note that in the vacuum the Hartree tadpole contribution is necessarily zero, but is in general finite in the medium). If we treat the damping parameter $\gamma_m$ in our phenomenological formula (\ref{eq:chi_theory}) for $\chi(\omega)$ as an adjustable parameter, then the interaction line (which is just a photon moving in the cavity, in interaction with the medium as well as the cavity) in the formula for $\Sigma ({\bf p}, \epsilon)$ also must be treated as phenomenological. On the other hand we can  calculate the photon propagator microscopically, in the framework of the same RPA. The photon propagator has the Dyson equation ${\cal D} = {\cal D}_o + {\cal D}_o \Pi_{HFA} {\cal D}$, where $\Pi_{HFA}({\bf q}, \omega)$ is the appropriate photon polarization part, incorporating, at RPA level, the effects of the medium and the cavity (compare Fig. \ref{fig:RPA}(d)).

\begin{figure}
\centering
\includegraphics[scale=0.35]{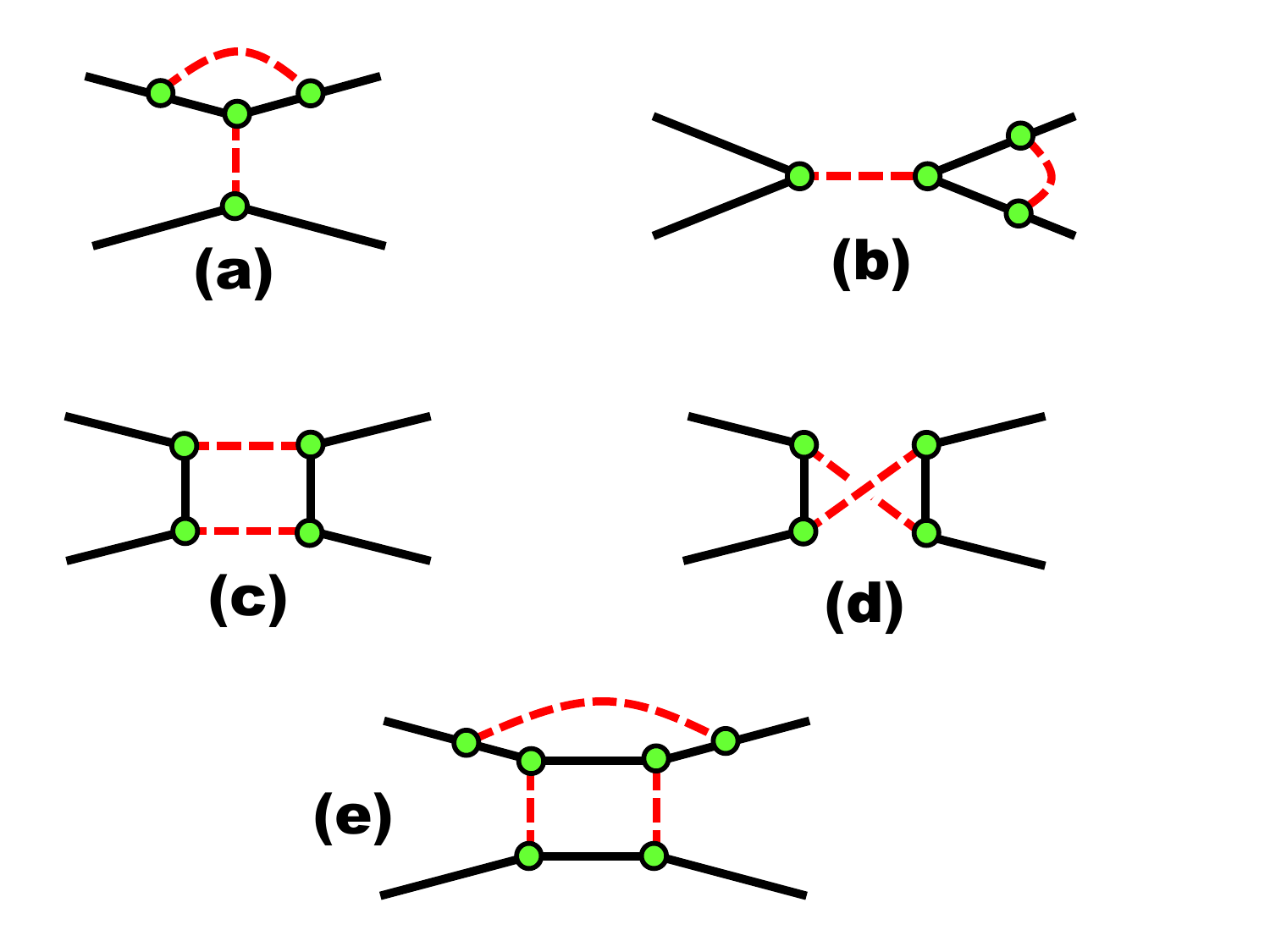}
\vspace{-7mm}
\caption{  Examples of corrections to the irreducible 4-point vertex, which play a role in determining the IR form of $\chi(\omega)$ when the internal photon momentum-energies go to zero. In (a) and (b) we depict 2nd-order vertex corrections to the lowest-order Hartree contribution; in (c), (d) we see two other 2nd-order corrections; and in (e) we see a 3rd-order term which provides a vertex correction to (c).        }
 \label{fig:RPA-corr}
\end{figure}

Without going into any details about such a calculation, we make the essential point here that such a calculation does {\bf not} include all of the contributions from those IR photons in interaction with the zero mode. Figs. \ref{fig:RPA-corr}(a) and \ref{fig:RPA-corr}(b) show contributions that are not included in the RPA irreducible 4-point vertex, but which will alter the result for this interaction, and these are just the first terms in an infinite set of contributions. To sum all of these to get the leading divergent behavior in the zero-energy limit, and how this affects the susceptibility and the damping function, is a formidable task (of considerably greater complexity than that involved in working all this out for the vacuum!), which we will not attempt here.

However, it is clear that if we correct the RPA by incorporating these extra IR terms, we will get an extra dissipative contribution to the response functions in the low-energy limit. The most immediate effect of this is the replacement of the superradiant phase transition---predicted by RPA---by the weakly avoided level crossing seen in the experiments. It will be of considerable interest, for future work, to give a complete quantitative theory of the photonic IR divergences for this system.

\subsubsection{The Role of Phonons}

Given the importance of soft photons for the understanding of the soft electronuclear mode, a natural question is what role do soft {\it phonons} play around the quantum critical point? As far as we are aware, there has been no attempt to investigate this question at all, a curious omission given that the spin-phonon couplings in the LiHoF$_4$ system are not weak \cite{bertaina06} and one sees clear phonon bottleneck behaviour in the spin dynamics.

Again, one can attempt to treat this problem in a conserving approximation, and many of the arguments go through as for the case of electronuclear interactions with photons. Thus, the structure of the diagrams for the dynamics susceptibility is the same as that shown in Figs. \ref{fig:RPA} and \ref{fig:RPA-corr}, with the photon lines replaced by phonon lines. One can even repeat the usual arguments which lead to the possibility of superradiance, this time mediated by phonons rather than photons.

However, again one is faced with the fact that the RPA conserving approximation does not capture all of the leading IR contributions to the 3-point spin-phonon vertex, or to the phonon-mediated spin-spin interactions. Analogous arguments to those used for photons indicate that vertex corrections will lead to extra sources of dissipation around the critical point in describing the soft mode propagator.

We are thus confronted with a situation in which two separate gapless bosonic fields couple simultaneously to the electronuclear soft mode. This is a formidable problem. The two IR contributions will not act independently of each other and there will be interference between the two, which will become pronounced at the approach to the quantum critical point. Even a simple model calculation is a challenging task, and we will not attempt it here, but reserve discussion to another paper. Note that there are multiple phonon branches in the LiHoF$_4$ system \cite{bertaina06}, and the electronuclear soft mode will couple to all of them.

{\it Summary}: There is a hierarchy of IR divergences in the interaction of photons with electron spin, with cancellations going on between these which lead to finite physical results at low energy. The RPA does incorporate some cancellation of IR divergences, but it does not give the complete story because conserving approximations like the RPA do not capture all the divergent terms, even at leading order. In cases where one has strong IR divergences (as in, e.g., the Kondo problem), this becomes a serious problem, and radically alters the low-T behavior. In the present case, the RPA appears to capture some key aspects of the behavior, at least for the coupling strengths prevailing in the experiment; however, it misses contributions to the dissipation coming from soft photons, which have the crucial effect of converting a singular response, with a predicted superradiant transition, to a weakly avoided level crossing. Fits to the data are possible, but we note that the full story of the IR divergent contributions is still to be resolved. In particular, understanding the manner in which spectral weight transfer enters as a variable is important to include in the interpretation of the fits in the RPA approximation. Although we have not attempted any kind of quantitative discussion, we fully expect that low-energy phonons also will modify the low-energy physics, via their coupling to the electronuclear soft mode.


\section{Experimental Results -- Soft mode spectra}
\label{sec:Spectra}

\begin{figure}
    \centering
    \includegraphics[width=2.75in]{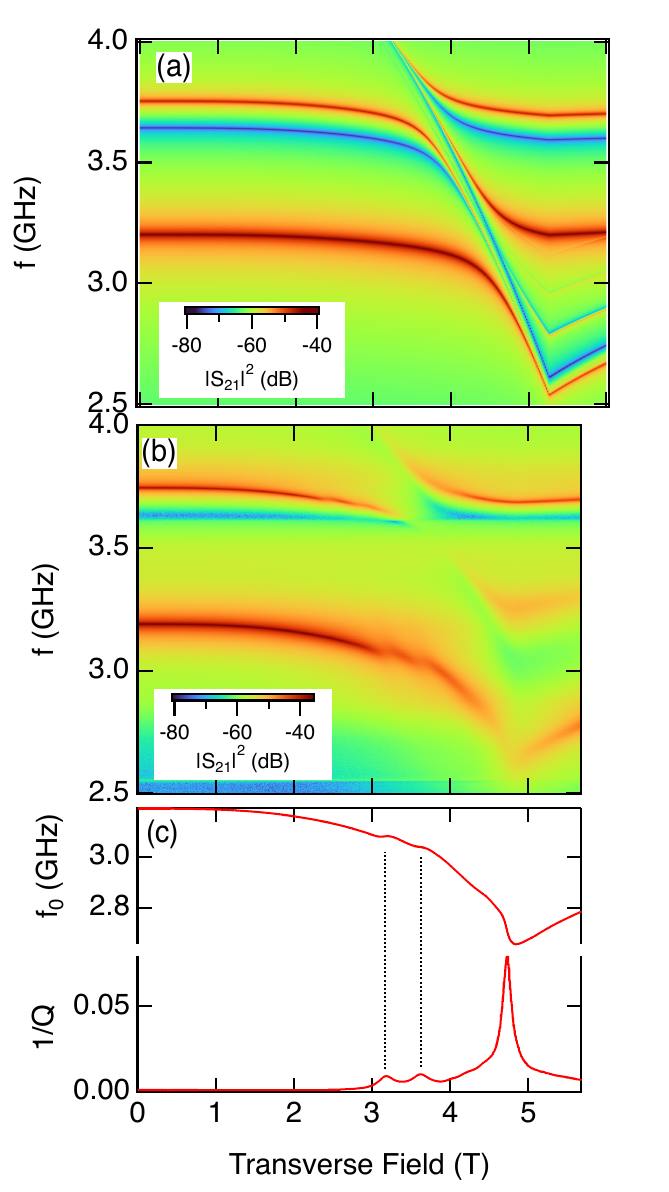}
    \caption{(a) Transmission spectrum $|S_{21}|^2$ for a bimodal resonator using the susceptibility calculated in the single-ion (MF) approximation input into the lumped element model. (b) Measured transmission of \LiHoF\ at $T=55$~mK corresponding to the calculation in (a). Strong coupling to the first excited state transition is visible, as well as avoided level crossings with the two cavity modes. Visible to the left of the dominant mode are a pair of smaller avoided level crossings due to a Walker mode and the soft collective mode, discussed in the text. (c) Results of fitting the cavity resonance starting at 3.2 GHz in (b) to a Lorentzian model. The center frequency $f_0$ (top) and inverse quality factor $1/Q$ (bottom), are related to the real and imaginary components of the dynamic susceptibility, respectively. Using the $1/Q$ data, we can fit another Lorentzian form to the small amplitude dual peaks, tracking a Walker mode and the collective soft mode discussed in the text. Finally, there is strong absorption just prior to the transition due to a level crossing with the lowest single-ion excitation. }
    \label{fig:highfreq}
\end{figure}

In order to connect the microscopic theory with the measured microwave spectra, we start with the dynamic susceptibility calculated using either a single-ion (MF) model or the RPA model described in Section \ref{sec:ENmodes} and then combine with the response of a resonator. If the cavity modes are weakly coupled, one may consider a linear superposition of the transmission from the two modes,
$S_{21}=S_{21}^a+S_{21}^b$, as discussed in the final paragraph of Section \ref{sec:resonator}). Under that same assumption, the absorption and dispersion of the sample can be extracted by fitting each $|S_{21}(f)|^2$ spectrum to a Lorentzian form:
\begin{equation}
    |S_{21}(f)|^2 =  \frac{A}{1+4 Q^2\left(f/f_0 -1\right)^2},
\end{equation}
with $f_0$ the center frequency and $Q$ the quality factor. $f_0$ serves as a measure of the real part of the sample susceptibility and $1/Q$ the imaginary part. For a bimodal resonator (see Fig.~\ref{fig:res} and the end of Section \ref{sec:resonator}), a comparison between model and measurement is shown in Fig.~\ref{fig:highfreq}. The MF model provides an accurate description of the majority of the features in the data. The dark band at 3.5 GHz is an indication of an anti-resonance in the LGR, the frequency of which is sensitive to the relative amplitudes of the two modes.  The principal exceptions are the avoided level crossings visible in the data at approximately 3.5 T and 3.1 GHz; these are ascribed to a magnetostatic (Walker) mode and a collective RPA soft mode.  We plot the results of these fits to the data in Fig.~\ref{fig:highfreq}(c). Both the dominant single ion excitation, and the weaker soft mode and Walker mode, are clearly visible as peaks in the dissipation.

\begin{figure}
    \centering
    \includegraphics[width=3.4in]{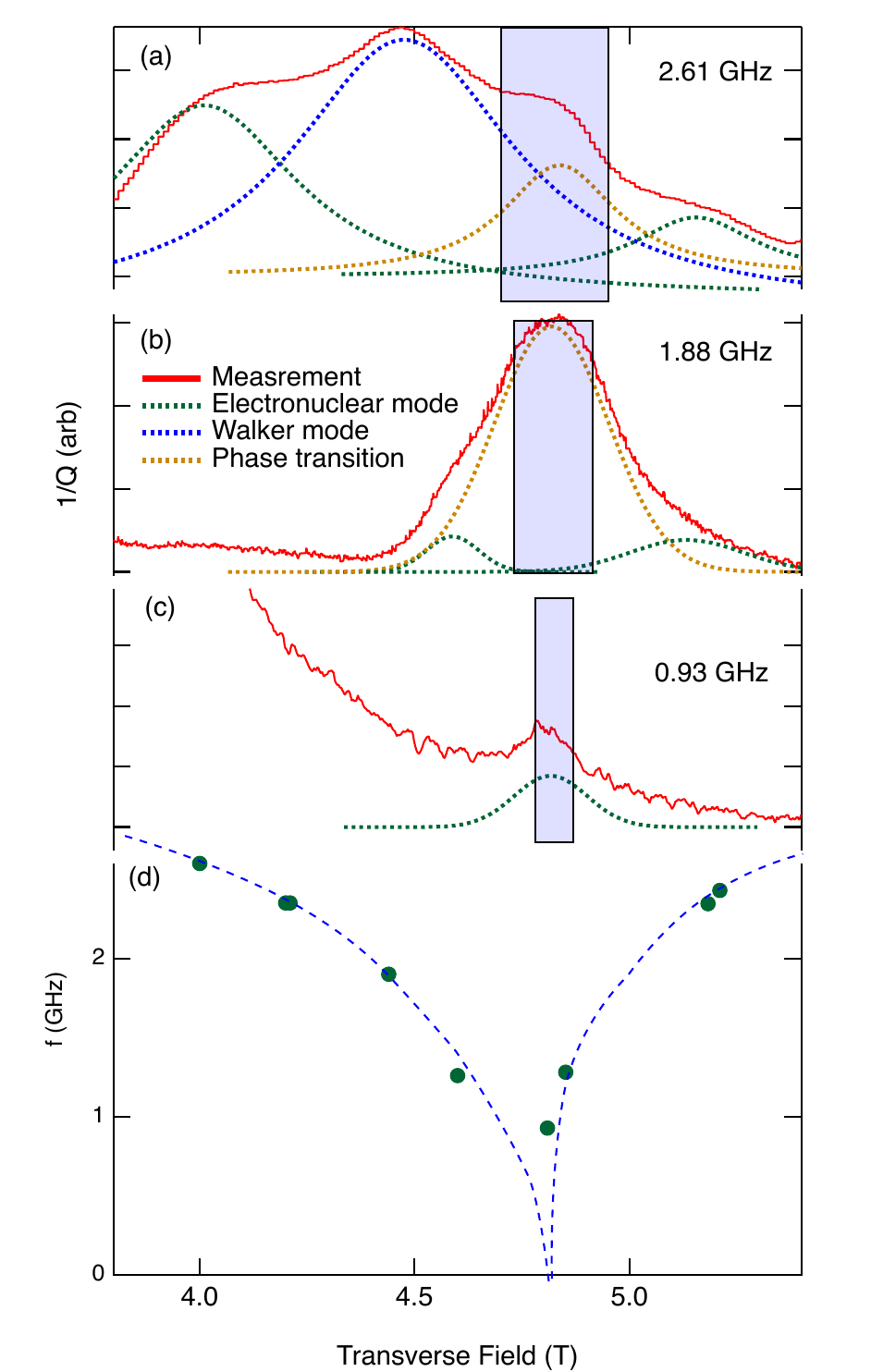}
    \caption{Mode structure in the vicinity of the quantum critical point. (a)-(c) Dissipation, $1/Q$, vs. transverse magnetic field for resonators tuned to 2.61, 1.88, and 0.93 GHz, respectively. Peaks in the dissipation arising from the electronuclear soft mode, magnetostatic Walker modes, and the quantum phase transition coexist, and can be fitted to a sum of Lorentzians (dashed curves). Shaded rectangles mark the approximate position of the quantum critical regime, with the center determined from the peak in the real susceptibility for each frequency and width equal to the field scale corresponding to $h\nu$ for each measurement. (d) Evolution of the electronuclear mode as a function of frequency; curve is a guide to the eye. The data shown in this panel were presented previously in Ref.~\cite{LiberskySM}.
    }
    \label{fig:softsummary}
\end{figure}

To track the evolution of the modes as a function of transverse magnetic field and frequency, we repeat these measurements for a series of resonator tunings, with the lower mode ranging from 0.9 to 2.9 GHz and the upper mode from 3 to 4.5 GHz. In the vicinity of the QPT, the excitations in LiHoF$_4$ primarily lie in the range spanned by the lower resonator mode; representative traces of $1/Q$ are shown in Fig.~\ref{fig:softsummary}. As the individual modes have finite width and in the vicinity of the phase transition often overlap, we perform simultaneous multi-mode fits to identify and track the different excitations (dashed lines in Fig.~\ref{fig:softsummary}). The assignment of experimental peaks in $1/Q$ to different underlying physical mechanisms is based on comparison with the single-ion and RPA models discussed above, the longitudinal magnetic-field dependence of the peaks \cite{LiberskySM} and a delineation of magnetostatic (Walker) modes discussed below in Section~\ref{sec:domains}. In particular, Fig.~\ref{fig:softsummary}(d) shows the evolution of the lowest-lying hybrid electronuclear mode, with the salient feature being the softening of the mode to energies below $k_B T$ in the vicinity of the QPT (adapted from Ref.~\cite{LiberskySM}).

\begin{figure}
    \includegraphics[width=3.4in]{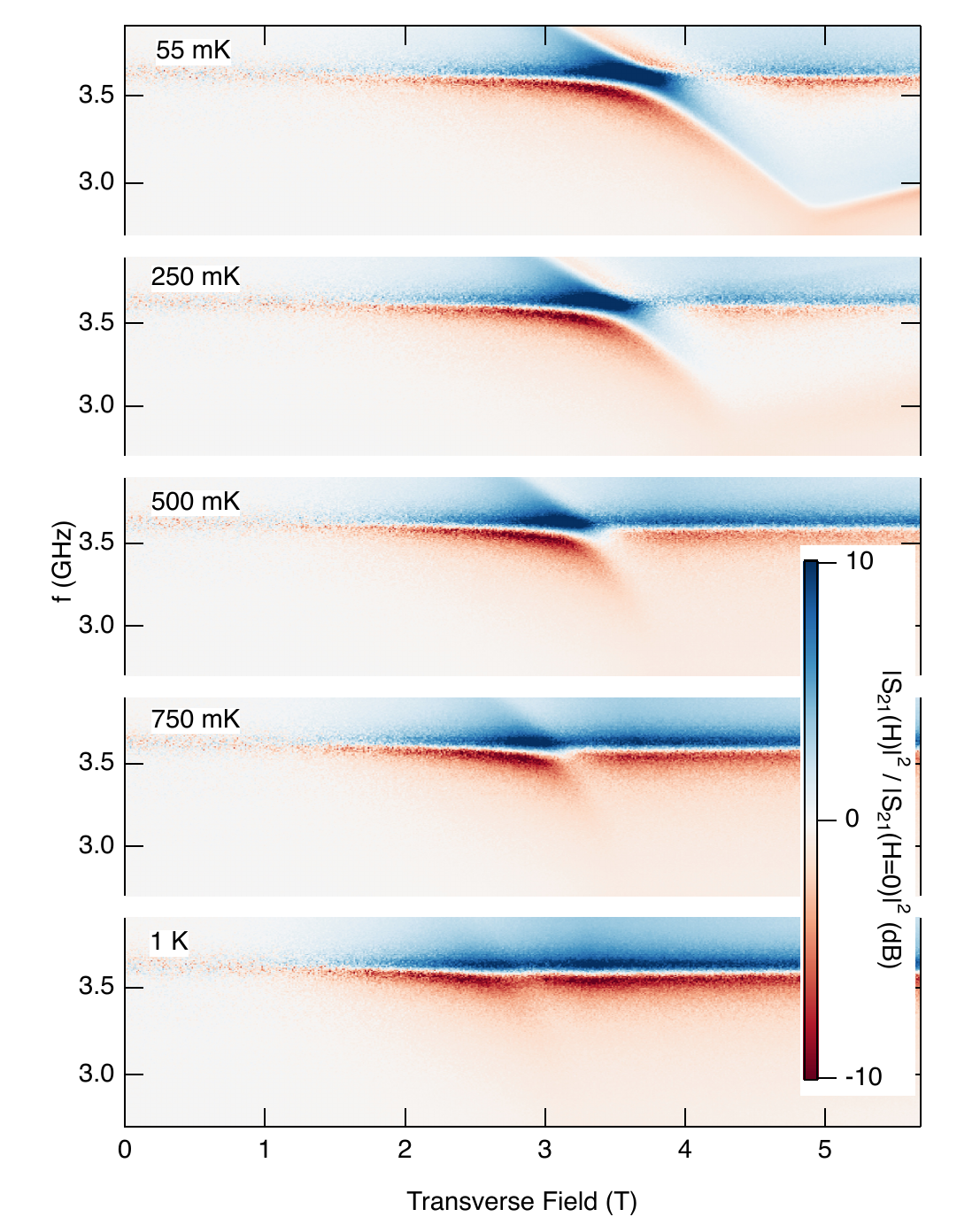}
	\caption{Temperature dependence of the off-resonance transition to the higher energy levels. Data are shown normalized by the $H_t=0$ transmission intensity. As the thermal energy becomes comparable to and greater than the mode splitting, the transitions wash out. Simultaneously, the transverse field scale decreases due to the shape of the ferromagnet to paramagnet phase boundary.}
	\label{fig:temp}
\end{figure}

We examine in Fig.~\ref{fig:temp} the temperature dependence of the microwave absorption to further characterize the behavior of the higher-energy transitions. Due to the relatively weak intensity of the off-resonance response, the data in this figure is shown normalized by the $H_t=0$ behavior. The two features dominant at low temperature---the cusp in the vicinity of the QPT and the avoided level crossing where the electronuclear mode energy approaches the 3.6 GHz upper resonance of the LGR---both diminish in intensity and move to lower transverse field  as the temperature is increased. The cusp associated with the paramagnetic transition is visible at 250 mK, but it is thermally washed out at higher $T$, while the avoided level crossing remains observable (albeit with lessened strength) to $T\sim1$ K, reflecting a separation of energy scales between single ion and collective behavior.

\begin{figure}
    \includegraphics[width=3.5in]{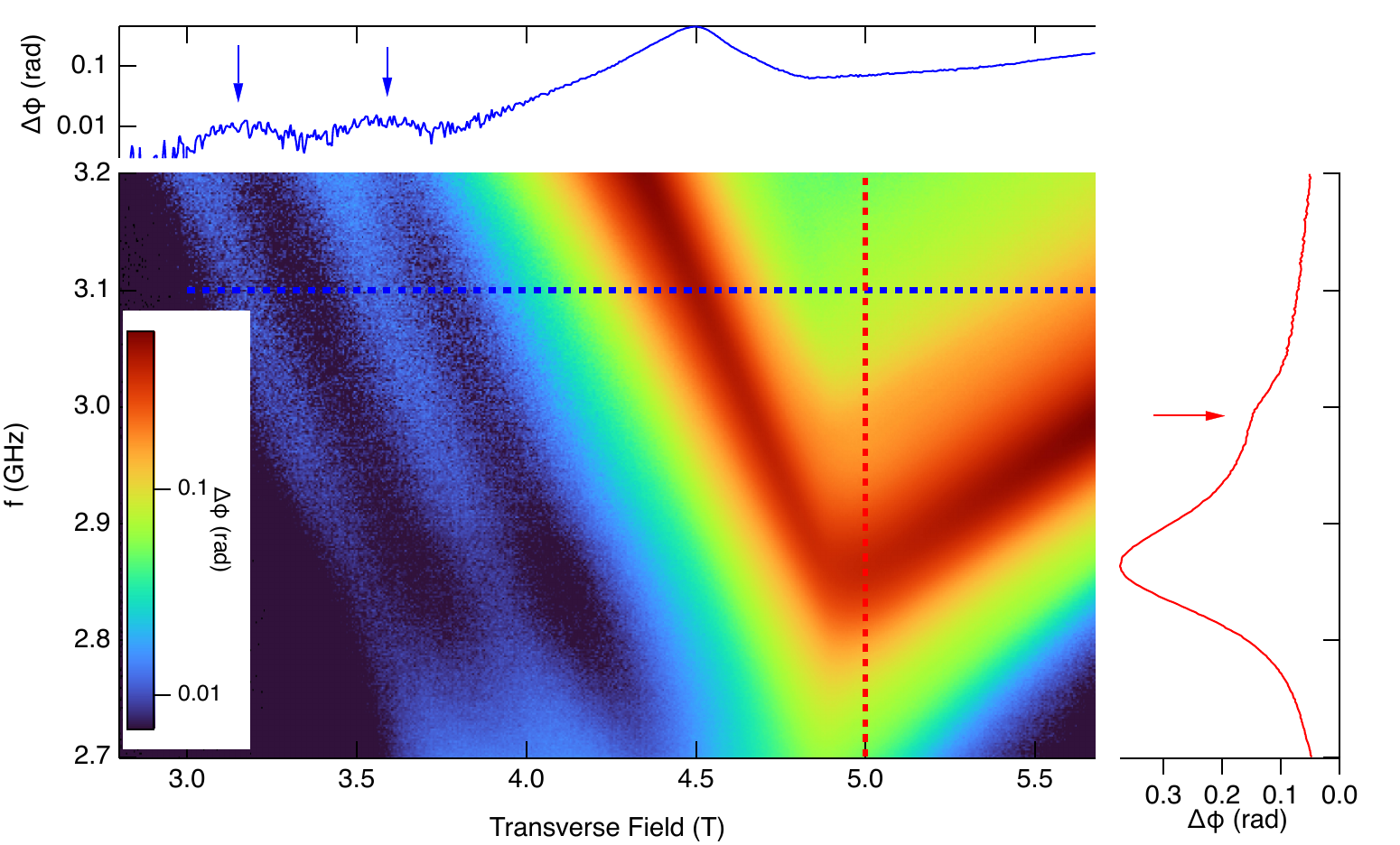}
	\caption{(a) Phase shift of $S_{21}$ relative to low-field behavior. The top panel shows a horizontal slice at 3.1 GHz; the dominant peak at 4.5 T is the lowest single-ion (MF) transition and the two small peaks marked with arrows indicate a Walker mode and the soft collective mode. The vertical slice at 5 T (right) shows the lowest single-ion transition (dominant peak) and a higher lying transition between excited-states (marked with arrow).
    }
	\label{fig:disperse}
\end{figure}

If the coupling between the microwave field and the sample is sufficiently strong, we can measure the dispersive response in the weak transmitted field regime away from the cavity resonance (at 4.2 GHz). This response is most easily seen in the relative phase shift of $S_{21}$, as illustrated in Fig.~\ref{fig:disperse}. Here, the frequency-dependent relative phase is defined relative to the behavior at low transverse field, $\Delta\phi(f,H)=\phi(f,H)-\langle\phi(f,H<3\:\mathrm{T})\rangle$.

Several features are apparent from the data in Fig.~\ref{fig:disperse}. The strongest feature corresponds to the lowest single-ion (MF) transition. It evolves continuously as a function of transverse field and has a sharp minimum at the ferromagnetic to paramagnetic phase transition. In the ferromagnetic phase, two features are visible below the dominant mode: a magnetostatic Walker mode and the collective soft mode. Finally, constant-field cuts reveal an additional mode above the dominant mode which persists into the paramagnetic phase. As such, it cannot arise from domain-based phenomena, and we identify it with a transition between excited states of the material.

\begin{figure}
    \includegraphics[width=3in]{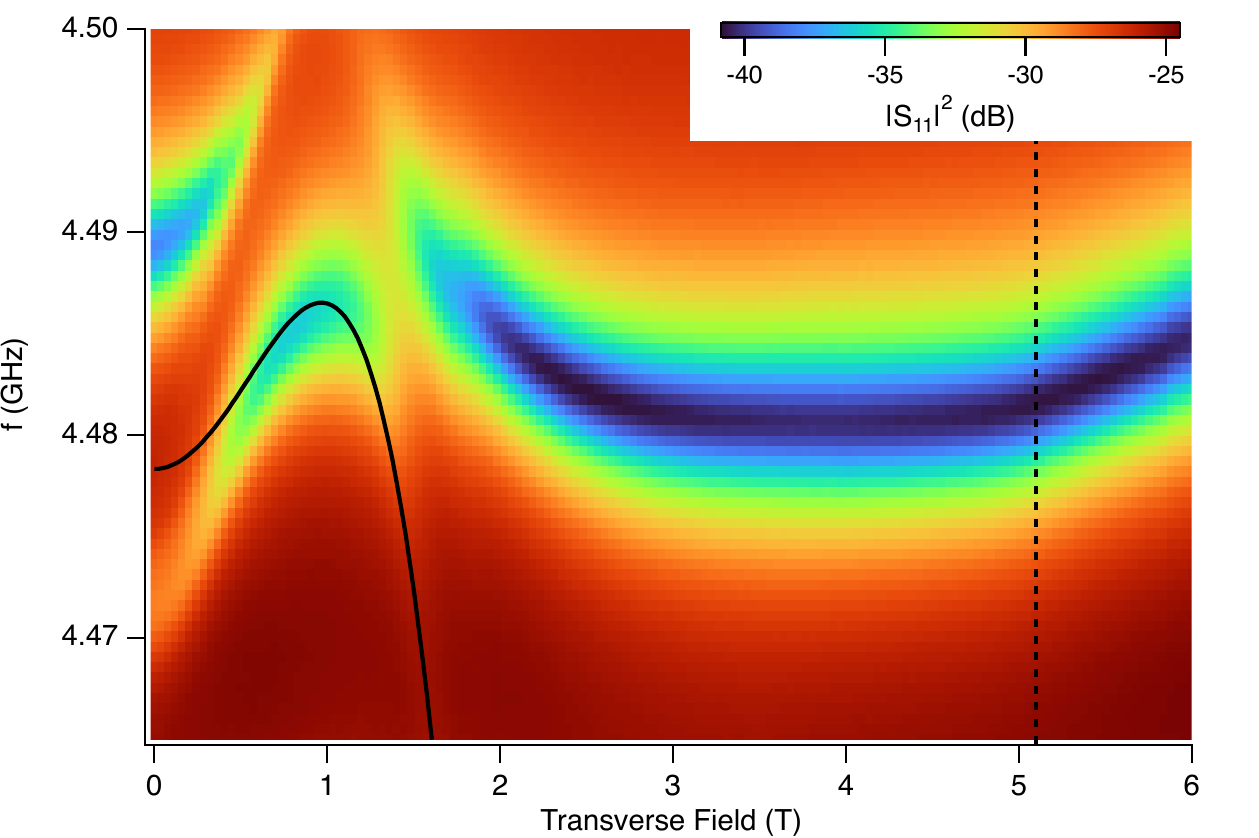}
	\caption{Transverse susceptibility $\chi^{yy}$. In the vicinity of the quantum phase transition (vertical dashed line), the RPA model predicts $\chi^{yy}\rightarrow 0$, consistent with the extremely-small measured sample response. At low transverse fields, a pair of avoided level crossings are observed between the resonator mode at 4.49 GHz and the non-monotonic evolution of the soft mode. The black curve shows the qualitatively-correct but approximately 0.15\% underestimate from the RPA prediction for that mode. The transverse field for the RPA calculation has been scaled up by 7\% to account for the field mismatch discussed in Ref.~\cite{ronnowMagneticExcitationsQuantum2007} and in Section~\ref{sec:resonator} above. }
	\label{fig:transverse}
\end{figure}

It is possible to test an additional aspect of the theory  \cite{mckenzieThermodynamicsQuantumIsing2018} by varying the orientation of the crystal's Ising axis with regard to the applied ac and dc magnetic fields. The  model predicts that in the vicinity of the quantum phase transition, the lowest-energy susceptibility for ac fields parallel to the crystalline Ising axis ($\chi^{zz}(\omega)$) diverges for the soft mode, whereas for ac fields in the transverse plane, $\chi^{yy}(\omega)$ vanishes at the quantum critical point. We test this prediction by reorienting the sample in the resonator such that the Ising axis (crystal $c$), the dc magnetic field, and the ac magnetic field of the dominant microwave mode are mutually perpendicular, thus probing $\chi^{yy}$. We show in Fig.~\ref{fig:transverse} that the resonator frequency varies by only 0.2\% across the entire field range, more than an order of magnitude smaller than the $\chi^{zz}$ response plotted in Figs. ~\ref{fig:lowfreq} and \ref{fig:highfreq}.  This experimental result is consistent with the RPA determination of a strongly-suppressed $\chi^{yy}$. Moreover, we are able to test the RPA calculation that the soft mode is non-monotonic as a function of transverse field, peaking at 4.487~GHz at approximately 0.9~T (solid curve in Fig.~\ref{fig:transverse}). This field scale is where the transverse field starts to mix hyperfine states of the system. With the resonator tuned to 4.49~GHz, it is possible to discern the pair of avoided level crossings seen in Fig.~\ref{fig:transverse} at transverse fields of 0.5 and 1.3 T. We note that the RPA predictions are a good quantitative match to the observed data, underestimating the energy by only 0.15\%.

\section{Effects of Longitudinal Field}
\label{sec:long_field}


Given that the soft mode is gapped by any finite longitudinal field, it is important to investigate its effect both theoretically and experimentally. In our cuboidal crystal, the multiple domain structure actually makes the longitudinal field more homogenous than it would be if there was a global demagnetization field coming from a single domain. This helps with the theoretical interpretation of the data.

In this Section, we first make some theoretical observations, and then look at the experiments using two different field ramping protocols.

\subsection{Theoretical Remarks}

We start by considering the effect of an inhomogeneous longitudinal field on the soft mode. We assume a longitudinal field component $B_z({\bf r})$ of the total field ${\bf B}({\bf r})$, varying with position ${\bf r}$ around the sample, and we introduce a ``density of states" function $N_z(\xi)$ for the longitudinal fields as a function of longitudinal bias $\xi$, defined as
\begin{equation}
N_z(\xi) \;=\; \int_{V_o} d^3 r \delta(\xi - B_z({\bf r}))
 \label{Nx-def}
\end{equation}
where we integrate over the entire volume $V_o$ of the sample. The number of spins which have a local longitudinal field in the range $\xi < B_z < \xi + d\xi$ is then $N_z(\xi) d\xi$.

It is useful to write $N_z(\xi)$ as
\begin{equation}
N_z(\xi) \;=\; \int dB_z |\nabla_{\bf r} B_z({\bf r})|^{-1} \delta (\xi - B_z)
 \label{Nz-xi}
\end{equation}
We then can imagine surfaces $\Sigma_{\xi}({\bf r})$ of constant $\xi$ in the sample, defined by the $\delta$-function in Eqn. (\ref{Nz-xi}). Their shape, and the way they move through a single-domain sample as we change $\xi$, is governed by Morse theory, and the basic ideas are similar to those used to discuss the density of states of phonons in solids. Hence we expect Morse/van Hove singularities as the surfaces pass in or out of the single-domain sample at specific values $\xi_j$, and an interesting structure for $N_z(\xi)$ that will depend on sample shape \cite{PS98,prokofevQuantumRelaxationEnsembles1998,villain99,tupitsyn04,takahashi11}.

The gradient function $\nabla_{\bf r} B_z({\bf r})$ is along a direction $\hat{n}_{\xi}({\bf r})$ perpendicular to the surface $\Sigma_{\xi}({\bf r})$, and $N_z(\xi)$ will increase as $|\nabla_{\bf r} B_z({\bf r})|$ decreases, diverging when $|\nabla_{\bf r} B_z({\bf r})| \rightarrow 0$. Any discontinuity in the area of the surface as a function of $\xi$ will lead to a discontinuity in $N_z(\xi)$; this happens when surfaces leave or enter the sample.

In thinking about this intuitively two tactics are useful. One is to look at solvable examples. For example, it is easy to work out the field distribution around a single-domain cylinder, and to plot the surfaces of constant $\xi$.

The plots of surfaces of constant $\xi$ give a second useful way of thinking about the physics. In any sample, the field varies in different parts of the sample because of inhomogeneous demagnetization fields - this is particularly true for any single-domain sample if its shape is far from ellipsoidal. If we now vary a constant applied longitudinal field $B^0_z$ to the sample, the surface $\xi = 0$ defines a set of points where the internal demagnetization field exactly cancels the external field, leaving net zero field. On this surface (which moves around in the sample as we vary $B_z$) the soft mode frequency will be identically zero. However, any departure from this surface will gap the soft mode. 

The effect of a spatially varying field in the sample, and hence of a spatially varying ${\bf B}_z({\bf r})$, is to then cause the energies of the collective modes in the sample to vary with position, insofar as their frequencies depend on $B_z$. This dependence is acute in the case of the zero mode in the \LiHoF\ system, in the region around the critical surface $\xi = 0$, because the soft mode frequency has a singular dependence on $B_z$ around $B_z = 0$ (see below).

Suppose we ignore the linewidth of the zero mode, and assume that it has a frequency $\omega_o(B_z, B_{\perp})$. We then can write the lineshape of the zero mode, as seen in the experiments, as
\begin{eqnarray}
S(\Omega) &=& \int d\xi \int_{V_o} d^3r \, \delta (\xi - B_z({\bf r})) \, \delta
\left(\Omega - \omega_o (\xi, B_{\perp}) \right) \nonumber \\
&=& \int dB_z \, N_z(B_z) \, \delta \left(\Omega - \omega_o(B_z, B_{\perp})
\right)
 \label{S-Om}
\end{eqnarray}

To proceed further we need a form for $\omega_o(B_z, B_{\perp})$. We posit an analytic form that fits the numerical results for $\omega_o(B_z, B_{\perp})$; these results were given in our earlier work \cite{LiberskySM}, and we find that a  good fit in the limit of low soft mode energy is provided as follows.

First we define:
\begin{eqnarray}
x^6 &=& \alpha \left( B_{\perp} - B_{\perp}^{min}(B_z) \right)  \nonumber \\
y^6 &=& \gamma B_z   \nonumber \\
z &=& x + iy
 \label{def-xyz}
\end{eqnarray}
where $\alpha, \gamma$ are constants, and $B_{\perp}^{min}(B_z)$ is the transverse field for which the zero mode energy is a minimum for a given longitudinal field $B_z$. If one looks at the numerical plots, one sees that $B_{\perp}^{min}(B_z) \sim O(B_z)^{1/2}$.

We now write a model form for the zero mode frequency as
\begin{eqnarray}
 \omega_o(z, \bar{z}) &=& \left(\alpha^{1/3} |B_{\perp} -
 B_{\perp}^{min}(B_z)|^{1/3} \;+\; \gamma^{1/3} |B_z|^{1/3} \right) \nonumber \\
 &=& z \bar{z}.
 \label{om-zz}
\end{eqnarray}
We note that the minimum value of $\omega_o$ when the transverse field $B_{\perp} = B_{\perp}^{min}(B_z)$ is $\propto B_z^{1/2}$. When $B_z = 0$, one has $B_{\perp}^{min}(0) = B^{\perp}_c$, the field at the quantum critical point.

If we now write the zero mode lineshape $S(\Omega, B_z, B_{\perp})$ as
\begin{equation}
S(\Omega) = \int d\omega_o \, \left({d \omega_o \over d \xi} \right)^{-1} N_z
(\omega_o) \, \delta(\Omega - \omega_o)
 \label{S-omw}
\end{equation}
we find that:
\begin{eqnarray}
S(\Omega) &=& \tfrac{3}{\gamma} \int d\omega_o \, N_z(\omega_o) \, \delta(\Omega -
\omega_o)   \nonumber \\
&& \qquad \times \; \left( \omega_o - \alpha^{1/3} |B_{\perp} -
B_{\perp}^{min}(B_z)| \right)^2
 \label{S-Om-f}
\end{eqnarray}
with a quadratic dependence on the frequency $\omega_o$. We see that this spectral weight picks up a very small contribution from the region where $\omega_o$ is small, i.e., the critical region where $B_{\perp} \sim H_c$ and where $B_z \sim 0$. This is a real problem in principle for experiments, because it is precisely the region where $\omega_o$ is small that is of greatest interest.

To see what happens analytically, suppose that we have a density of states function that is flat and can be written as
\begin{equation}
N_z(\xi) \;=\; {1 \over B_2 - B_1} \left[ \theta(B_2 - \xi) - \theta(B_1 - \xi)
\right]
 \label{square}
\end{equation}
so that we then have
\begin{eqnarray}
S(\Omega) &=& \tfrac{3}{\gamma (B_2-B_1)} \int^{\omega_o(B_2)}_{\omega_o(B_1)}
d\omega_o  \, \delta(\Omega - \omega_o) \nonumber \\
&& \qquad \times \; \left( \omega_o - \alpha^{1/3} |B_{\perp} -
B_{\perp}^{min}(B_z)| \right)^2
 \label{S-Om-sq}
\end{eqnarray}

The quadratic dependence on $\omega_o$ in the integrand now strongly emphasizes the regions of the sample where $|\omega_o|$ is large, i.e., in the wings where $|B_z|$ is the largest of $|B_2|$ or $|B_1|$. The region where $\omega_o$ is small is hardly captured at all.

It is thus clear that to see the soft mode in the region where it really goes soft, we require a very small spread in longitudinal fields in the sample. More detailed numerical fits of $\omega_o(B_z, B_{\perp})$, which are accurate outside the very low energy regime, indicate that we need to have a spread $\Delta B_z$ in longitudinal fields which is less than $\sin 200$ Oe in order to see the soft mode. As discussed previously \cite{LiberskySM}, this appears to the be the case in our samples precisely because we have multiple domains, which then smooth the field $B_z({\bf r})$ around the sample.

A more detailed analysis would use, instead of (\ref{om-zz}) for the zero mode frequency $\omega_o$, a form for the dynamic susceptibility $\chi(\Omega, \omega_o; \xi, B_{\perp})$ of the zero mode; and we would use the correct distribution function $N_z(\xi)$ for the longitudinal fields in the sample, computed for what we believe is the correct domain structure in the sample. The resulting lineshape predicted for the experiments would then be
\begin{equation}
S(\Omega, B_{\perp}) \;=\; - \int {d\xi \over \pi} \, N_z(\xi) \, Im \chi(\Omega, \omega_o; \xi, B_{\perp})
 \label{S-chi}
\end{equation}

A more thorough investigation of this question requires characterization of the actual field distribution in the sample. Nonetheless, the fact that the existence of multiple domains in the sample smooths the demagnetization field \cite{LiberskySM} helps enormously in seeing a relatively sharply-defined mode in the experiments (see also discussion in section \ref{sec:domains} below).

\subsection{Experimental Results}

\begin{figure*}[ht]
    \includegraphics[width=6.5in]{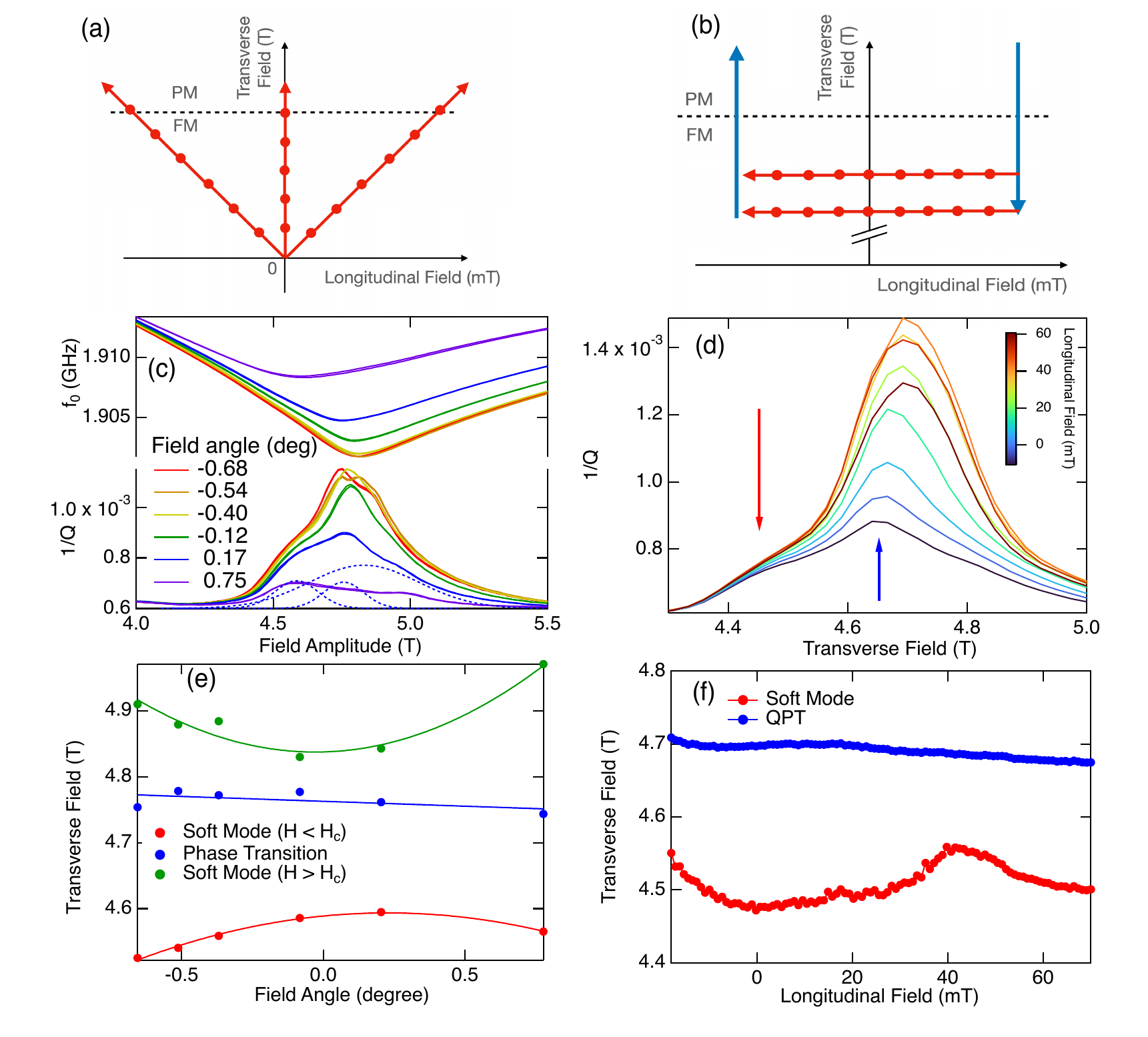}	
	\caption{ Effects of longitudinal dc field. (a) Fixed-angle protocol: fields along the transverse and longitudinal directions are ramped together ratiometrically.  (b) Field-cool protocol: starting in the paramagnet, a longitudinal pinning field is applied, the transverse field is decreased to the measurement field, and data taken as a function of longitudinal field. (c) Response to the fixed-angle protocol. The center frequency (top) shows the shift in the quantum phase transition cusp as the field angle is changed. The absorption (1/Q) (bottom) of the low-energy mode absorption diminishes for larger field angles. Zero degrees denotes a field purely in the transverse direction with respect to the crystal. Dashed lines show the three fitted peaks for the $0.17\deg$ trace. (d) Response to the field-cooling protocol, showing a large peak in the dissipation at the quantum phase transition and a smaller satellite peak due to the low-energy mode crossing the 1.9 GHz measurement frequency. (e) Field-angle dependence of the low-energy mode for transverse fields above and below the quantum phase transition, along with the angle dependence of the phase transition itself. Curves are guides to the eye. (f) Loci of the phase transition and low-energy mode in $(H_l,H_T)$ space. For both field protocols, the low-energy mode exhibits qualitatively different behavior than the phase transition, revealing the influence of a dc longitudinal field and the onset of history dependence due to ferromagnetic domain formation.}
	\label{fig:long}
\end{figure*}

We now look at what is actually seen. Applying a dc magnetic field along the Ising axis provides insight into magnetic domain formation and the relationship of ferromagnetic domains to collective mode dynamics. Due to the potentially hysteretic nature of the ferromagnet, two distinct field ramping protocols were investigated. First, as depicted in  the right column of Fig.~\ref{fig:long}, the direction of the field vector is fixed and the amplitude ramped continuously from 0 to 6 T. Multiple overlapping peaks in the dissipation are observed (dashed curves in Fig.~\ref{fig:long}(c)). In order to classify these peaks, we plot their dependence on field angle in Fig.~\ref{fig:long}(e). The response of the low-energy soft mode is qualitatively different to that of the quantum phase transition, allowing the two features to be separated. As expected, the QPT (identified by the minimum in $f_0$, corresponding to the peak in the static susceptibility) depends only weakly on the longitudinal field, whereas the enhanced dissipation associated with the soft mode has a strong and approximately quadratic dependence on applied field (Fig.~\ref{fig:long}(e)). This dependence is seen for the mode both above and below the quantum critical point.

A second field-sweeping protocol was  followed (right column of Fig.~\ref{fig:long}) to control for ferromagnetic domain effects. In this protocol, we start at high transverse field, sufficient to place the sample well inside the quantum paramagnetic phase. A longitudinal pinning field of 70 mT is applied, the transverse field is ramped down to the desired value immediately below the critical field, thereby driving the sample into the ferromagnetic state. The response of the \LiHoF\ sample is then measured as a function of longitudinal field as that field is ramped through zero to the maximum accessible negative field for our coil set. This process is repeated for a series of transverse fields, defining a grid in $(H_l,H_t)$ parameter space. Cutting through the grid at a series of constant longitudinal fields yields the curves shown in Fig.~\ref{fig:long}(d), where we now see the phase transition and the soft-mode feature in the ferromagnetic state, but the initial pinning field has suppressed the soft-mode feature previously visible in the paramagnetic phase. As shown in Fig.~\ref{fig:long}(e), the features defining the phase transition and  the soft mode respond to longitudinal field in qualitatively different fashions. The transverse-field location of the phase transition only depends weakly on the longitudinal field. However, the presence of a longitudinal field does suppress the magnitude of the peak in the dissipation. By contrast, the location of the soft-mode dissipative peak moves significantly as a function of longitudinal field. It is asymmetric about zero, indicating that hysteresis due to ferromagnetic domains plays a significant role in setting the field scale for the mode.


\section{Domain structure and Walker Modes}
\label{sec:domains}

Previous measurements have examined the domain structure in \lhf\ in the classical regime by means of optical measurements \cite{battisonFerromagnetismLithiumHolmium1975, pommierPHASEDIAGRAMUNIAXIAL1988} and scanning Hall probe microscopy \cite{eisenlohrFrontiersQuantumCriticality}. Stripe domains form oriented along the Ising axis, with characteristic lateral scale of 2 microns \cite{pommierPHASEDIAGRAMUNIAXIAL1988}. As the results reported here entailed using a cuboidal sample, we also must consider the non-uniform demagnetizing field distribution. This aspect of the physics has been discussed previously in the context of \lhf \cite{Biltmo:2009ck,twengstromLiHoFCuboidalDemagnetizing2020,mckenzieMagnetostaticModesCriticality2023}. The demagnetizing fields act to broaden the observed spin resonances, and in an applied ac field they are responsible for the spatially inhomogeneous Walker modes discussed below.

Up to this point, our calculations of the dynamic susceptibility have assumed uniform or negligible demagnetization fields in a sample where the magnetization is uniform locally. This assumption includes uniformly magnetized ellipsoids, and spins within a particular domain (away from the domain walls) of an arbitrarily shaped sample where neighboring domains eliminate the local demagnetization field. Modes present in such a system, known as Kittel modes \cite{kittelTheoryFerromagneticResonance1948}, contain uniformly precessing spin excitations. In the presence of a spatially non-uniform ac field and/or non-uniform magnetization or demagnetization fields in the sample (due to domains, the sample shape, etc.), modes which have a spatially-varying phase within the sample can be excited. These are known as magnetostatic or Walker modes, after the description by Walker~\cite{walkerMagnetostaticModesFerromagnetic1957}. The term ``magnetostatic'' is a misnomer given that these are intrinsically dynamic phenomena; the term refers to the fact that the amplitude of the time dependent field and magnetization fluctuations are assumed to obey the classical magnetostatic equations. 

In the measurement, Walker modes are visible for a wide range of transverse fields, as shown in Fig.~\ref{fig:disperse}.  While such modes in general can be excited by inhomogeneities in either the rf field \cite{walkerMagnetostaticModesFerromagnetic1957}, or the magnetization and demagnetization field fluctuations, here the high homogeneity of the rf field provided by the loop-gap resonator design \cite{liberskyDesignLoopgapResonator2019a} suggests that it is the latter inhomogeneity which is principally responsible for enabling the Walker mode behavior.

Several factors are at play in the analysis of such modes in this system, including the coupling between the electronic and nuclear magnetization, and the domain structure. To account for the domain structure, we consider a sample composed of equal and opposite magnetic stripes in the $xz$ plane, with the magnetization $\boldsymbol{M} = (m_x,0,\pm m_z)$. To account for the coupled electronuclear modes, we consider the equations of motion of the set of electronic and nuclear spin operators $ \{ \boldsymbol{X}_i \} = \{ \boldsymbol{\tau}_i, \boldsymbol{I}_i \}$, as in equations (\ref{eq:motion}) and (\ref{eq:dynamics}).

The effect of stripe domains has been examined in the context of other magnets~\cite{Kittel1946, polderResonancePhenomenaFerrites1953, KooyEnz,sigalFerromagneticResonanceAbsorption1979}. The magnetization on the top surface of the sample is $\sigma_m=m_z \text{sgn}[\cos{(\pi y/d)}]$, where $d$ is the width of the stripes. We expand the surface magnetization in a Fourier cosine series so that the magnetization is symmetric about $y=0$ with the first domain wall occurring at $y=\pm d/2$. Following Kooy and Enz \cite{KooyEnz}, we solve for the magnetostatic potential inside and outside the sample. The resulting demagnetization field inside the sample is given by
\begin{align}
\label{eq:demag}
H_D^z(y,z) &= - \sum_{n'} \frac{4m_z}{n \pi} e^{-\frac{\pi n L_z}{2d}}
\cos{\biggr(\frac{\pi n y}{d}\biggr)} \cosh{\biggr(\frac{\pi n z}{d} \biggr)}
\\ \nonumber
H_D^y(y,z) &= - \sum_{n'} \frac{4m_z}{n \pi} e^{-\frac{\pi n L_z}{2d}}
\sin{\biggr(\frac{\pi n y}{d}\biggr)} \sinh{\biggr(\frac{\pi n z}{d} \biggr)},
\end{align}
where $n' = \{n : n>0, n \ \text{odd}\}$. We assume a sample with a finite length $L_z$ that extends to infinity in the transverse directions. Note that in a system comprised of many stripe domains, for which $L_z/d >> 1$, the demagnetization field is suppressed exponentially. The width of the domains follows from minimization of the magnetostatic free energy of the system, taking into account the energy of the domain walls. This solution of the magnetostatic equations neglects the crystal field anisotropy present in LiHoF$_4$. In order to account for anisotropy, we must make contact between the microscopic spin Hamiltonian given in equation (\ref{eq:LiHoTrunc}), and the macroscopic field and magnetization given by the magnetostatic equations.

The demagnetization fields given in equation (\ref{eq:demag}) are a solution to the magnetostatic equations, $\nabla \times \boldsymbol{H}=0$ and $\nabla \cdot \boldsymbol{H} = -\nabla \cdot \boldsymbol{M}$. To make contact with the microscopic spin Hamiltonian, we consider the local field acting on each holmium ion, given by $\textbf{H}_{loc} = \textbf{H}_a + \textbf{H}_{dip} + \textbf{H}_{ex} + \textbf{H}_{hyp}$, where we consider an externally applied field $\textbf{H}_a$, a field due to the dipolar interactions $\textbf{H}_{dip}$, the exchange field $\textbf{H}_{ex}$, and the hyperfine field due to the nuclear spins $\textbf{H}_{hyp}$. We analyze each term in this expression below.

The local moment at each lattice site is the sum of the electronic and nuclear moments of each holmium ion, $\boldsymbol{M}= \boldsymbol{m}^e+\boldsymbol{m}^n = \rho_s \gamma_J \langle \boldsymbol{J}_i \rangle + \rho_s\gamma_I \langle \boldsymbol{I}_i \rangle$, where $\rho_s$ is the spin density, and respectively, $\gamma_J \langle \boldsymbol{J}_i \rangle$ and $\gamma_I \langle \boldsymbol{I}_i \rangle$ are the electronic and nuclear moments at site $i$. In LiHoF$_4$ there are four spins per unit cell with volume $V_{cell}=a^2 c$, where the transverse lattice spacing is $a=5.175${\AA}, and the longitudinal lattice spacing is $c=10.75${\AA}. This gives a spin density of $\rho_s = 4/V_{cell} =  1.3894 \times 10^{28} \text{m}^{-3}$. The nuclear moment is small ($|\gamma_I/\gamma_J| \ll 1$), so that the magnetization at each lattice site is $\boldsymbol{M} \approx \boldsymbol{m}^e = -\rho_s g \mu_B \langle \boldsymbol{J}_i \rangle$, where $g=5/4$ is the Land{\'e} g factor, and $\mu_B$ is the Bohr magneton. Although the nuclear contribution to the magnetization is small, one must consider both the electronic and nuclear spin contributions to the magnetization dynamics; this is due to the strong hyperfine coupling present in LiHoF$_4$.

The substantial hyperfine coupling in LiHoF$_4$ leads to a strong hyperfine field acting on the electronic spins. In the equation of motion, when the electronic and nuclear spins are decoupled in the RPA, this leads to a strong local nuclear field acting on each electronic spin. This approximation is problematic near the quantum critical point of the system, where electronuclear correlations become important. The Green's function approach to calculating the electronuclear modes avoids this difficulty. Away from the critical point, where electronuclear correlations are less important, one may perform an RPA decoupling of the electronic and nuclear spins. Given the slow dynamics of the nuclear spins, the hyperfine field is then an additional static field acting on the electronic spins.

The dipolar and exchange fields stem from the interaction in equation (\ref{eq:LiHoTrunc}), $V_{ij}^{zz} = J_D D_{ij}^{zz} - J_{nn}$. The antiferromagnetic exchange field is $H_{ex}^z = \lambda_{ex} M_z$, with $\lambda_{ex}=-2.73 \times 10^{-2}$ (see the supplement of reference \cite{LiberskySM}). The zero wavevector component of the dipole sum is
\begin{align}
D_0^{zz}  =  \rho_s \biggr[\frac{4\pi}{3} + \lambda_{dip} - 4\pi N_z\biggr],
\end{align}
where $\lambda_{dip}=1.664$ is the lattice correction and $N_z$ is the demagnetization factor. In general, the local dipolar field acting on the spins is \cite{AharoniBook}
\begin{align}
\label{eq:Hdip}
\textbf{H}_{dip} = \frac{1}{4\pi}
\biggr[\frac{4\pi}{3}\textbf{M} + \textbf{H}_{\Lambda} \biggr] +
\boldsymbol{H}_D,
\end{align}
where the terms in square brackets are the Lorentz local field and a lattice correction, and the final term is the demagnetization field. In LiHoF$_4$, we assume that the crystal field quenches all but the longitudinal component of the dipolar field. The lattice correction is then $H_{\Lambda}^z = \lambda_{dip} M_z$. In a uniformly magnetized ellipsoid $H_D^z = -N_z M_z$, where $N_z$ is the demagnetization factor. To account for the domain structure, we introduce an effective demagnetization factor so that $H_D^z = -N_z^{eff} m_z$, with $m_z$ being the magnitude of the local magnetization inside a particular domain \cite{LiberskySM}  (for further details of demagnetization fields in cuboidal samples see \cite{twengstromLiHoFCuboidalDemagnetizing2020}). In the absence of an applied field the domain structure arranges itself so the demagnetization field is zero and $N_z^{eff} \approx 0$. This validates the use of a needle-shaped sample in calculations..

The demagnetization field due to a stripe domain pattern, given in (\ref{eq:demag}), contains a $y$ component which is quenched by the crystal field in LiHoF$_4$, assuming the surface spins are subject to the same crystal field as the bulk spins. In the experiment, with a static field applied in the $x$ direction, there will be magnetic surface charge on the $yz$ surfaces of the sample. The resulting demagnetization fields are also assumed to be quenched or vanish due to the crystal symmetry. We leave corrections to these approximations, and the demagnetization fields resulting from the corners and edges of the sample, as subjects for future work.

\begin{figure}[tb]
    \centering
    \includegraphics[width=3in]{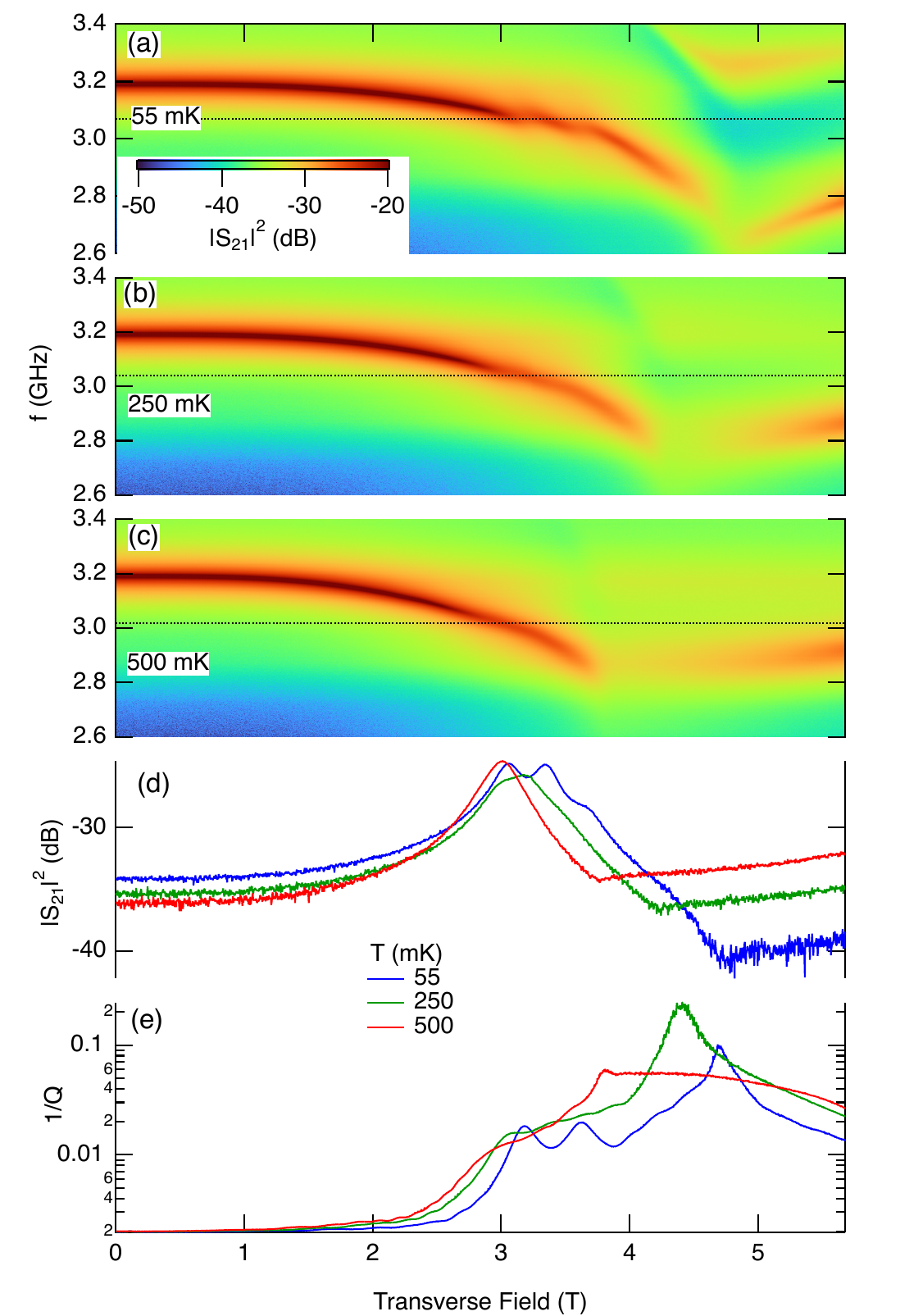}
    \caption{Temperature evolution of Walker modes, measured with a resonator tuned to 3.2 GHz at zero magnetic field. (a-c) As the temperature increases from 55 to 500 mK, the avoided level crossings characteristic of the Walker modes weaken and vanish, while the behavior driven by the electronuclear mode structure remains robust. (d) constant-frequency cuts (shown by the dashed lines in the upper three panels) exhibit the temperature evolution. The cuts evolve from a clear multi-peak structure to a single peak due to the net magnetization of the sample with no additional features. (e) $1/Q$ vs $H_T$ for three temperatures, showing the thermal suppression of the Walker mode features.}
    \label{fig:TWalker}
\end{figure}

The applied field, exchange field, hyperfine field, and the component of the dipolar field given in square brackets in equation (\ref{eq:Hdip}), are homogeneous over short length scales so that their curl and divergence are zero. The magnetostatic equations are then $\nabla \times \boldsymbol{H}_D=0$ and $\nabla \cdot \boldsymbol{H}_D = -\nabla \cdot \boldsymbol{M}$, where the demagnetization field and the magnetization of the sample may be inhomogeneous. This connects the microscopic spin Hamiltonian with the magnetization and field determined by Maxwell's equations. We return now to the LiHoF$_4$ system in an applied ac field, and determine the Walker modes present in the system.

Consider a ferromagnetic phase of the system, in which the stripe domain pattern is present. In the experiment, the uniform ac field leads to a uniform magnetic fluctuation across the cuboidal sample. The resulting non-uniform demagnetization field fluctuation stirs up a Walker mode. The zero wavevector component of this Walker mode excitation shows up in the experiment. In order to calculate the average demagnetization field within a stripe in a multidomain sample, one may average the demagnetization field present in a single stripe of the stripe domain phase. This leads to an alternating series with terms that decay like $1/r^3$, as one would expect if the terms correspond to demagnetization corrections from neighboring stripe domains at an increasing distance from the stripe of interest. Assuming this to be the case, we truncate the series at the first term to obtain the average demagnetization field with $z$ and $x$ dimensions $L_z$ and $d_{w}$.

Averaging the longitudinal demagnetization field in equation (\ref{eq:demag}) over a single stripe, one finds that
\begin{align}
\label{eq:HDavg}
\overline{H}_D^z &= \frac{1}{L_z L_y d_w}
\int_{-d_w/2}^{d_w/2} \int_{-L_y/2}^{L_y/2} \int_{-L_z/2}^{L_z/2} H_D^z \\ \nonumber
&= - M_z \frac{8 d_w}{L_z} \sum_{n'} \frac{\sin{(\pi n/2)} }{(n \pi)^3}
\biggr[1-\exp{\biggr(-\frac{\pi n L_z}{d_w}\biggr)}\biggr],
\end{align}
where the summation is over odd positive values of $n$. The dimensions of the LiHoF$_4$ sample are $V_{sample} = L_x \times L_y \times L_z =$ 1.8 mm $\times$ 2.5 mm $\times$ 2.0 mm, so that $L_z/L_x = 1.11$. The longitudinal magnetization is $M^z = - \rho_s g \mu_B C_{zz} \langle \tau^z \rangle$, with $\langle J^z \rangle = C_{zz} \langle \tau^z \rangle$ being the truncated longitudinal spin operator. In zero transverse field, the truncation parameter is $C_{zz} = 5.51$, which leads to a magnetization given by $\mu_0 M^z = - 1.115 \langle \tau^z \rangle$ T. Setting $L_x=d_w$, and truncating the alternating series in equation (\ref{eq:HDavg}) at the first term, one finds the average demagnetization field within a stripe in a multidomain sample to be $\mu_0 \overline{H}_D^z = 0.245 \langle \tau^z \rangle$ T.

We look at the temperature dependence of the two types of modes to further separate out the domain dynamics manifesting as Walker modes from the soft mode and other excitations driven by the single-ion energy hierarchy of the sample. As shown in Fig.~\ref{fig:TWalker}, the avoided level crossings associated with the Walker modes are clearly visible at 55 mK, but have essentially vanished by 500 mK. Conversely, transitions to the higher electronuclear modes persist to 1 K and above (Fig.~\ref{fig:temp}).


\section{Conclusions}
\label{sec:Conclusion}


The pure LiHoF$_4$ ferromagnet features experimentally-accessible electronic and nuclear spin degrees of freedom in a theoretically tractable package, with the further dimension of controllable disorder provided by the ability to substitute magnetic Ho ions with non-magnetic Y. In this paper we have explored what may be the most striking aspect of the onset of long-range magnetic order: the quantum critical point that is posited to exist for a quantum Ising magnet in a strong transverse field. At first glance, there is no obvious reason why there should be a simple quantum critical point in LiHoF$_4$, simply because the delocalized electronic magnons couple via strong hyperfine interactions to a large set of localized nuclear excitations. Previous experimental and theoretical work on systems of this kind \cite{gatteschi,takahashi11,Prokofev:2000bd} suggested that the nuclear spins should act to destructively scatter any coherent spin waves.

However, as first shown \cite{mckenzieThermodynamicsQuantumIsing2018} in 2018, the actual picture when the hyperfine couplings are strong should be one in which the nuclear spins hybridize with the electronic spins to form a set of delocalized electronuclear modes. These electronuclear modes include a single mode which softens around the quantum critical point, along with a large number of gapped modes, as shown in Fig. \ref{fig:energy}.

In the real world of LiHoF$_4$, things are more complicated than this simple picture. The electronuclear soft mode couples to two other sets of soft modes, viz., photons and phonons, which are both gapless. This coupling severely changes some of the physics, particularly if the coupling is strong. The resulting spectrum is not at all obvious. Remarkably, the electronuclear soft mode survives, but one finds theoretically that the intensity of this mode will be weak as energy goes to zero because of an infrared cancellation mechanism, reminiscent of the physics that characterizes the Kondo problem.

This is not the only reason that observation of the soft mode is difficult. Another is its extraordinary sensitivity to any finite longitudinal field, which very rapidly gaps the soft mode; and any inhomogeneity in the longitudinal fields will then smear out the soft mode completely, rendering it invisible. As discussed above, in any single-domain non-ellipsoidal sample, where the longitudinal fields vary across the sample, one expects this to happen. The only reason we do see the mode is because the multiple domains smooth the field at the length scale of the soft mode wavelength.

For these reasons, and others, the experimental observation of the hybridized photon-electronuclear magnon modes became a real challenge, with the first observation reported only a few years ago \cite{LiberskySM}. This then raises the inevitable question of what is still left to understand and explore, and what further surprises may be waiting. We should like to emphasize the following questions:

(i) {\it Disorder and impurities}: Defects and impurities are known to have a profound effect on quantum phase transitions \cite{millis01,varma02,Ji21}. What happens to the electronuclear modes, particularly the soft mode, on substitution of a small fraction of the Ho ions by Y (or some other species)? 

(ii) {\it Effect of Longitudinal Fields}: It would be of considerable interest to further explore the profound effect of longitudinal fields on the soft electronuclear mode. The results will depend very sensitively on the domain structure in the sample and on sample shape as well as the field-sweeping protocol.

(iii) {\it Infrared Divergences}: A full clarification of the role of IR divergences coming from photons interacting strongly with many-body systems in cavities is, in our view, still outstanding. Away from the quantum critical point the problem is more tractable, but in the present case we are dealing with the interplay between two different soft modes, each with its own IR peculiarities. In the real world we also have to deal with soft long wavelength acoustic phonons, which enter the fray as one approaches the quantum critical point. 

(iv) {\it Implications for Quantum Information Processing}: One of the reasons for the  enduring interest in quantum critical phenomena in the LiHoF$_4$ system is the link to adiabatic quantum computation \cite{lidar18}. It is well known that the coupling of the system to a ``spin bath'' of two-level systems must strongly affect the slow sweep dynamics through the critical point \cite{wan09,wild16}. However, the role of the soft mode has yet to be understood. In particular, how does the electronuclear soft mode interact with the two-level systems, and how does this affect the transition through the critical point? One can ask the same about any spin impurities in the system, which brings us back to the questions posed in (i) above.

\acknowledgements{The experimental work at Caltech was supported by the U.S. Department of Energy Basic Energy Sciences Award No. DE-SC0014866. Theoretical work at UBC was supported by the National Sciences and Engineering Research Council of Canada, grant No. RGPIN-2019-05582. We gratefully acknowledge C. Simon for useful discussions.}

%

\end{document}